\documentclass[fleqn,usenatbib]{mnras}

\usepackage{newtxtext,newtxmath}
\usepackage[T1]{fontenc}
\usepackage{ae,aecompl}

\usepackage{graphicx}	% Including figure files
\usepackage{amsmath}	% Advanced maths commands
\usepackage{amssymb}	% Extra maths symbols
\usepackage{todonotes}
\usepackage{times}
\usepackage{xcolor}
\usepackage[normalem]{ulem}

\newcommand*\Bell{\ensuremath{\boldsymbol\ell}}

\defcitealias{Gonzalez:sub}{Paper~I}

\usepackage{etoolbox}
\makeatletter
\patchcmd\@combinedblfloats{\box\@outputbox}{\unvbox\@outputbox}{}{\errmessage{\noexpand patch failed}}
\makeatother

\title[HD~100453 II -- The hidden companion]{Spirals, shadows \& precession in HD~100453 -- II. The hidden companion}
\author[R. Nealon et al.]{Rebecca Nealon$^{1}$\thanks{E-mail: rebecca.nealon@leicester.ac.uk},
Nicol\'as Cuello$^{2}$,
Jean-Fran\c{c}ois Gonzalez$^{3}$,
Gerrit van der Plas$^{2}$,
\newauthor
Christophe Pinte$^{4}$,
Richard Alexander$^{1}$,
Fran\c cois M\'enard$^{2}$
and Daniel J. Price$^{4}$
\\
$^{1}$Department of Physics and Astronomy, University of Leicester, University Road, Leicester LE1 7RH, UK\\
$^{2}$Univ. Grenoble Alpes, CNRS, IPAG (UMR 5274), F-38000 Grenoble, France\\
$^{3}$Univ Lyon, Univ Claude Bernard Lyon 1, ENS de Lyon, CNRS, Centre de Recherche Astrophysique de Lyon UMR5574, F-69230, Saint-Genis-Laval, France\\
$^{4}$School of Physics and Astronomy, Monash University, Clayton VIC 3800, Australia\\
}

\date{Accepted XXX. Received YYY; in original form ZZZ}

\pubyear{2020}

\begin{document}
\label{firstpage}
\pagerange{\pageref{firstpage}--\pageref{lastpage}}
\maketitle

\begin{abstract}
The protoplanetary disc HD~100453 exhibits a curious combination of spirals, shadows and a relative misalignment between the observed outer disc and inferred inner disc. This disc is accompanied by a secondary star on a bound orbit exterior to the disc. Recent observations have suggested there may be an additional low-mass companion residing within the disc inner cavity. In our companion paper the orbit of the secondary was shown  to be misaligned by $61\degr$ to the plane of the outer disc. Here we investigate the properties of the inner companion and the origin of the misalignment between the inner and outer disc. Using numerical simulations and synthetic observations, we show that the disc structure and kinematics are consistent with a $\lesssim5$~M$_{\rm J}$ planet located at $15-20$~au. We find that the disc evolution over $\sim50$ binary orbits ($\sim 10^5$~yrs) is governed by differential precession and to a lesser extent, the Kozai-Lidov effect. In our proposed model the misalignment observed between the outer and inner disc arises naturally as a result of the misaligned outer companion driving the outer disc to precess more rapidly than the inner disc.
\end{abstract} 


%%%%%%%%%%%%%%%%%%%%%%%%%%%%%%%%%%%%%%%%%%%%%%%%%%

%%%%%%%%%%%%%%%%% BODY OF PAPER %%%%%%%%%%%%%%%%%%

%\listoftodos

\section{Introduction}
\label{sec:intro}
Observations of protoplanetary discs have revealed a wealth of substructure including spiral arms, rings, gaps, misalignments and warps. Internal disc processes have been proposed to explain some of these features, including dust sintering \citep{Okuzumi:2016by}, snow surfaces \citep{Stammler:2017bw}, self-induced dust traps \citep{Gonzalez:2017bu}, magnetohydrodynamic effects \citep{Bethune:2016wf}, winds \citep{Riols:2019cw} and zonal flows \citep{Flock:2015vq}. These features may alternatively be generated by the interaction with companions such as gap sculpting planets \citep[e.g.][]{Dipierro:2015od,Ruiz:2016ne,Pinte:2020gh} and external companions \citep{Dong:2016oc,Cuello:2019bd,Cuello:2020rt,Menard:2020aa}. The protoplanetary disc around HD~100453 exhibits spiral arms \citep{Wagner:2015ph,Dong:2016oc}, narrow lane shadows \citep[likely from a misaligned inner disc,][]{Benisty:2017kq}, a dust cavity \citep{Wagner:2015ph}, and a bound binary companion \citep{Chen:2006vs,Collins:2009he}. Additionally, the inner and outer discs appear to be misaligned \citep{Benisty:2017kq}, there is a warp across the outer disc, and a misalignment between the outer disc and the companion \citep{vanderPlas:2019gy}. A complete picture of the HD~100453 system must thus simultaneously explain the observed disc features, the multiple planes of misalignment, and the influence of the exterior companion.

HD~100453~A is an A9Ve star with an age of $6.5$~Myr, a mass of $1.7\,$M$_\odot$ and an accretion rate of $1.4 \times 10^{-9} \, M_{\odot}/{\rm yr}$ \citep{Collins:2009he,Vioque:2018pg}. A companion star HD~100453~B was first identified by \citet{Chen:2006vs} and subsequently associated to the primary by \citet{Collins:2009he} with a mass of $0.2\pm0.04\,$M$_\odot$. The protoplanetary disc surrounding the primary extends between $\sim21-42$~au in the near-infrared and displays a two-armed spiral structure extending to $38$~au \citep{Wagner:2015ph}. \citet{vanderPlas:2019gy} measured the mass of the disc between $0.001$ and $0.003\,$M$_\odot$ using the CO isotopologue line ratios with a corresponding gas to dust ratio between 15--45 (with the uncertainty stemming from the $^{12}$CO/$^{18}$CO ratio). From continuum emission the disc is observed to have an inclination of $29.5\degr$ and a position angle of $151.0\degr$. Observations of $^{12}$CO, $^{13}$CO, C$^{18}$O J=2--1 emission lines by \citet{vanderPlas:2019gy} also found evidence of a warp across the outer disc of $\sim10 \degr$. The characteristics of the narrow lane shadows in scattered light suggest a misalignment between the inner and outer disc of $72\degr$ \citep{Benisty:2017kq,Min:2017oc}.

It has additionally been suggested that a companion resides between the inner and outer disc \citep{Wagner:2015ph,vanderPlas:2019gy,Rosotti:2020nj}. This `inner companion' would naturally explain the low mass accretion rate onto the primary and the observed dust cavity interior to 21~au \citep{Wagner:2015ph}. \citet{vanderPlas:2019gy} additionally suggested that this inner companion may be responsible for the strong misalignment between the inner and outer disc. In HD~100453, such a companion is likely to have a mass between $0.01\sim0.1\,$M$_\odot$ and be located around 13~au \citep{vanderPlas:2019gy}. As it has so far eluded detection in the kinematics, this inner companion is likely to be of planetary rather than stellar mass. This picture is consistent with a number of other discs that also display cavities where planetary or low stellar mass companions are thought to reside such as PDS~70 \citep{Keppler:2018pd,Mueller:2018vw,Keppler:2019ce}, AB~Aur \citep{Boccaletti:2020sp,Poblete:2020} and HD~142527 \citep{Marino:2015rh,Casassus:2015yu}.

The evolution of an inner companion, disc and outer companion becomes complex when the outer companion is misaligned. This general scenario has previously been investigated using numerical simulations by \citet{Xiang-Gruess:2014ne}, \citet{Martin:2014bb}, \citet{Picogna:2015bo} and \citet{Lubow:2016nw}. Both \citet{Martin:2016qf} and \citet{Picogna:2015bo} considered scenarios where the outer companion was inclined enough ($\gtrsim 39 \degr$) that the Kozai-Lidov mechanism \citep{Kozai:1962gq,Lidov:1962gq} was able to act on the disc and inner companion. The simulations by \citet{Picogna:2015bo} found that the inner companion was able to decouple from the disc because the perturbations from the outer companion dominate the damping by the disc. After the inner companion and disc decouple, the inner companion's evolution is well described by purely gravitational ($N$-body) dynamics. This behaviour was also found by \citet{Martin:2016qf}, where the orbit of the inner companion `circulated' (where the precession rate and inclination is independent of the outer disc) when the mass of the disc was less than that of the inner companion. In the context of HD~100453, the evolution of an additional disc located interior to the inner companion has not yet been considered. 

\citet{vanderPlas:2019gy} additionally suggested that the proposed inner companion could be responsible for the relative misalignment of the inner and outer disc. \citet{Owen:2017oj} showed that the excitation of a secular resonance between the inner disc and companion commonly results in relative misalignments of more than $60\degr$ within a few million years. \citet{Zhu:2018vf} also found it was possible to make large relative misalignments using an inclined companion residing in a disc using numerical simulations. However, both of these works made limiting assumptions that have been shown to alter the relative misalignment that can be achieved (\citealt{Owen:2017oj} neglected viscous damping effects and \citealt{Zhu:2018vf} fixed the planet orbit, preventing its inclination from damping). When these effects are taken into account, the maximum relative misalignment found is reduced \citep[e.g. taking into account planet migration and conserving angular momentum,][]{Xiang-Gruess:2013fg}. Finally, the proposed inner companion in HD~100453 must have a low mass to avoid detection in the existing kinematics \citep{vanderPlas:2019gy,Rosotti:2020nj}, likely less than a Jupiter mass. For a companion of this size, inclination and eccentricity damping is rapid \citep{Tanaka:2004od}. It is thus not clear that the proposed inner companion alone can cause the observed $72\degr$ misalignment between the inner and outer disc.

In our companion paper, \citet[][ henceforth Paper~I]{Gonzalez:sub} we establish that the most likely orbit of the outer companion is misaligned to the outer disc by $61\degr$. In this work we will investigate the properties of the inner companion and examine the long term evolution of the complete HD~100453 system. Section~\ref{section:recap} summarises the key observations, and the main findings from \citetalias{Gonzalez:sub} that apply here. In Section~\ref{section:planet} we use numerical simulations to infer the location and mass of the inner companion. In Section~\ref{section:long_term_evolution} we use $N$-body calculations to show the long term evolution, taking into account differential precession and the Kozai-Lidov effect. In doing so we will show that the relative misalignment between the inner and outer disc depends on the presence of the inner companion but is necessarily driven by the outer companion. We discuss our results in Section~\ref{section:disc}~and draw conclusions in Section~\ref{section:conc}.

\begin{figure}
    \centering
    \includegraphics[width=\columnwidth]{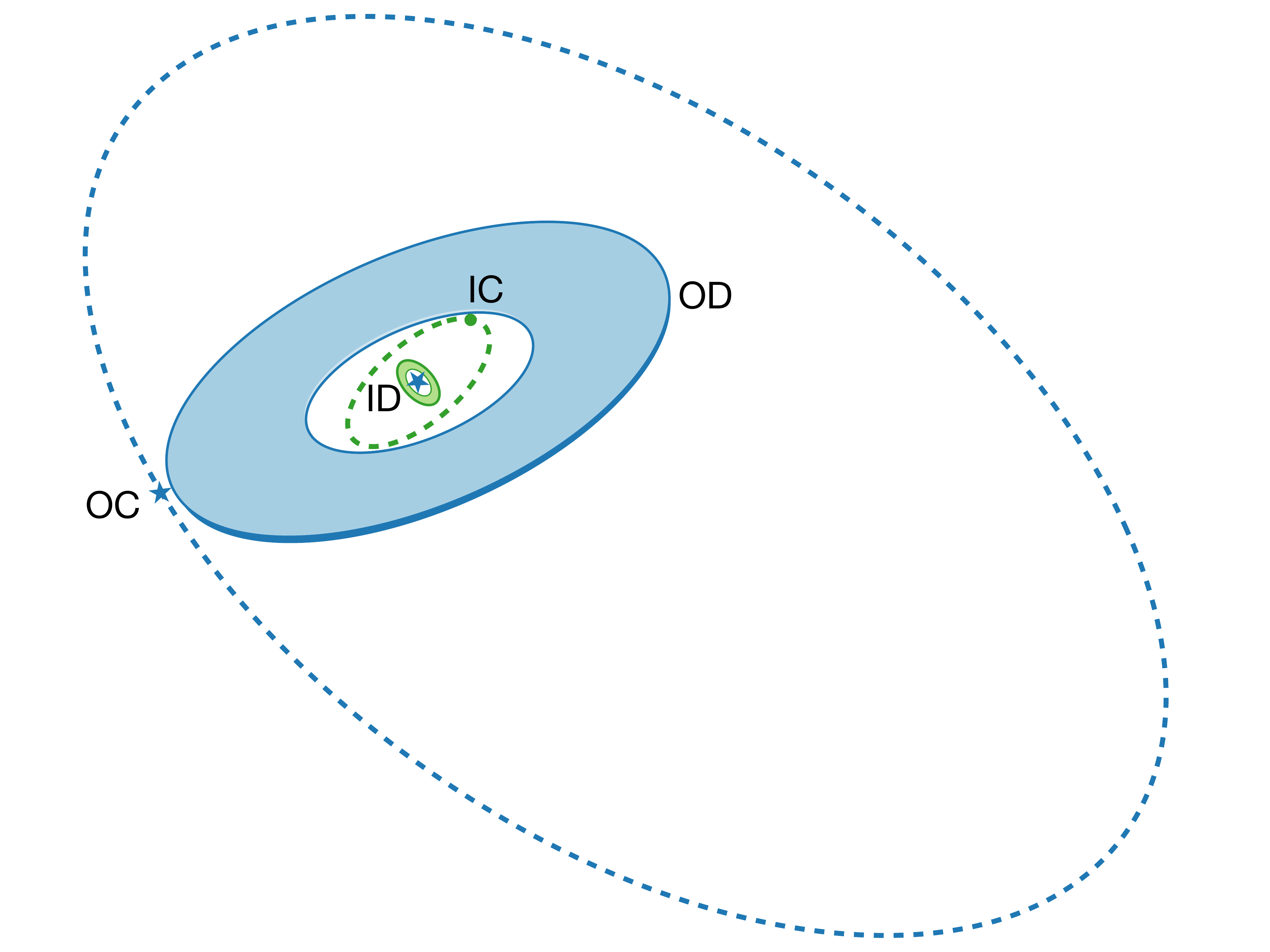}
    \caption{Not-to-scale schematic of the different components of HD~100453 system: The outer companion (OC, in blue), outer disc (OD, in blue), proposed inner companion (IC, in green) and inferred inner disc (ID, in green). This view is roughly in the plane of the sky. The relative misalignment between each component is expressed with the unit angular momentum vectors $\Bell$, where $\Bell_{\rm ID} \cdot \Bell_{\rm OD} \approx 72 \degr$ \citep{Benisty:2017kq}, $\Bell_{\rm OD} \cdot \Bell_{\rm OC} \approx 61 \degr$ (\citetalias{Gonzalez:sub}) and $\Bell_{\rm IC} \cdot \Bell_{\rm OD}$ is not required to be zero.}
    \label{fig:HD100-sketch}
\end{figure}

%-------------------------------------------------------------
%-------------------------------------------------------------
\section{Observational constraints}
\label{section:recap}
Here we summarise the constraints on the extent of the inner and outer disc, the relative misalignment between the two, the properties of the binary orbit and the findings we use from \citetalias{Gonzalez:sub}. The components of HD~100453 are summarised in Figure~\ref{fig:HD100-sketch}.

HD~100453's outer dust disc extends roughly from 20 to 40~au with a cavity at its inner edge. The inner edge of the outer disc was first measured at 21~au using scattered light \citep{Wagner:2015ph}. Subsequent observations using GPI polarised intensity imagery measured the outer disc between $18$--$39 \pm 2$~au \citep{Long:2018vt} and using mm dust between $23$--$40$~au \citep{vanderPlas:2019gy}. In Section~\ref{section:planet_properties} we require that the inner edge of the outer disc be truncated between $18$--$23$~au to be consistent with these measurements. The outer disc displays two symmetric, prominent spiral arms that are identified in both scattered light \citep{Wagner:2015ph,Benisty:2017kq} and in the CO emission \citep{Rosotti:2020nj}. These spirals were suggested to be due to tidal interaction with the outer companion \citep{Dong:2016oc}. The CO emission also suggests a warp of about $10\degr$ across the outer disc \citep{vanderPlas:2019gy}.

Constraints on the size of the inner disc are not as strong as for the outer disc. Observations in near-IR and mid-IR measure its half light radius around 1~au \citep{Menu:2015vy,Lazareff:2017ug} and thermal emission in H-band failed to detect an inner cavity \citep{Kluska:2020og}. Spectral energy distribution (SED) fitting by \citet{Long:2018vt} suggests that the inner disc extends between 0.13 and $1.0 \pm 0.5$~au, consistent with VLTI/MIDI estimates of $0.9\pm0.1$~au. Scattered light observations by \citet{Benisty:2017kq} found two narrow lane shadows cast across the outer disc, demonstrating a strong relative misalignment between the inner and outer disc. Following \citet{Min:2017oc}, modelling of the shadows suggest a relative misalignment of $72 \degr$, corroborated by \citet{Kluska:2020og}. \citet{Long:2018vt} suggest a smaller relative misalignment of $45 \pm 10 \degr$ to be consistent with their SED fitting and the separation of the shadows. For the inner disc, in this work we adopt an outer radius of 1~au, an inner radius of 0.1~au and a relative misalignment of $72\degr$.

The orbit of HD~100453~B (the outer companion) has been partially constrained using astrometric fits by both \citet{Wagner:2018hs} and \citet{vanderPlas:2019gy}. The former measured an orbit approximately co-planar with the outer disc with semi-major axis $a=1\farcs06\pm0\farcs09$, eccentricity $e=0.17\pm0.07$ and inclination $i=32\fdg5\pm6\fdg5$. Noting that the disc extended further than the Roche Lobe for a co-planar orbit, \citet{vanderPlas:2019gy} instead suggested that the orbit was likely misaligned to the outer disc. Using a Markov chain Monte Carlo approach they found that the orbit with the maximum likelihood had a relative inclination of $\Delta i=61 \degr$, where $\Delta i$ is measured with respect to the outer disc plane.

In our companion \citetalias{Gonzalez:sub} we demonstrated that the best fit orbit of \citet{vanderPlas:2019gy} is the most likely for HD~100453. We adopted this best fit orbit with $a=207$~au, $e=0.32$ and $\Delta i=61 \degr$ along with the three next best fitting orbits (ranked by $\chi^2$) and modelled the disc evolution using hydrodynamical simulations. The overall disc morphology, spiral features and velocity structure were all matched for the best fitting orbit and poorly by the other three best fits. In this work we thus adopt the binary parameters from \citetalias{Gonzalez:sub}, summarised in the top of Table~\ref{table:properties}.

%-------------------------------------------------------------
%-------------------------------------------------------------
\section{Dynamical hints of an inner companion}
\label{section:planet}
\subsection{Can HD~100453~B explain the broken inner disc?}
\label{section:analytical}
No. While \citet{Dogan:2018wo} demonstrate that the external torque provided by a stellar companion in a misaligned circumprimary disc can result in disc breaking, this is not the case for HD~100453. In the disc breaking scenario, each annulus of the disc experiences an individual torque from the binary due to its distance from the misaligned companion, resulting in differential precession of the disc. Disc `breaking' occurs when this differential precession results in the disc splitting into two distinct new discs \citep[e.g.][]{facchini_2013} while disc `tearing' occurs when the disc is torn into multiple independently precessing rings \citep[e.g.][]{nixon_2013}. As discussed by \citet{Dogan:2018wo}, disc breaking with an outer companion is an expected outcome when the disc communicates on a time-scale longer than the precession driven by the misaligned companion.

To test whether this scenario occurs for HD~100453 we compare the sound crossing and precession time-scales. Assuming a typical aspect ratio of $H/R = 0.05$ at 1~au, the sound crossing time-scale between the inner edge $R_{\rm in}$ and the outer edge $R_{\rm out}$ can be expressed as
\begin{align}
    T_{\rm s} = \int_{R_{\rm in}}^{R_{\rm out}} \frac{2}{c_{\rm s,0} (R/R_{\rm 0})^{-q}} \, {\rm d}R\,\,\,.
    \label{equation:ts}
\end{align}
Here $q$ determines the radial profile of the sound speed, $R_{\rm 0}$ is the reference radius and $c_{\rm s,0}$ is set by the aspect ratio at that reference radius. We use a vertically isothermal equation of state, $c_{\rm s}(R) = c_{\rm s,0} (R/R_{\rm 0})^{-q}$, with $q=0.25$. Between an assumed inner edge of $R_{\rm in} \sim 0.01$~au and the outer radius of $R_{\rm out} = 40$~au, this equates to roughly 1.5 outer binary orbits (one orbit of the outer binary takes $2161$~yrs).

Taking into account its eccentric orbit, the outer companion drives the outer disc to precess on a timescale given by (\citealt{Bate:2000fk}; \citetalias{Gonzalez:sub})
\begin{equation}
\frac{T_{\rm p}}{T_{\rm b}} = \frac{1}{K\cos {\Delta i}} \frac{\sqrt{1+q}}{q}\left(\frac{R_{\rm out}}{a(1-e_{\rm b}^2)}\right)^{-3/2}\,\,\,,
\label{equation:tp}
\end{equation}
where $T_\mathrm{b}$ is the period of the outer binary and
\begin{align}
K=\frac{3}{4}R_{\rm out}^{-3/2}\frac{\int_{R_{\rm in}}^{R_{\rm out}}\Sigma(r) r^3\,{\rm d}r}{\int_{R_{\rm in}}^{R_{\rm out}}\Sigma(r) r^{3/2}\,{\rm d}r}.
\label{Eq:K_Terquem}
\end{align}

Using a $\Sigma(R)$ profile with a taper at the inner edge, the properties listed in Table~\ref{table:properties} and $R_{\rm in} = 0.01$~au we estimate a single precession of the disc to be as rapid as $\sim 6.5\times10^5$~yrs. As the disc precesses on a much longer time-scale than the disc communicates the presence of a warp (i.e. $T_{\rm p} > T_{\rm s}$), the break in the disc cannot be caused by the outer companion. From this we conclude that there must be an as yet unobserved body residing in the gap of HD~100453 which acts to separate the disc into the inner and outer disc that are observed. This supports previous suggestions by \citet{Wagner:2015ph,vanderPlas:2019gy} and \citet{Rosotti:2020nj}.

\begin{table}
\caption{Parameters of HD~100453 system from \citetalias{Gonzalez:sub} used in this work.}
\label{table:properties}
\centering
\begin{tabular}{@{}lcr@{}}
\hline
Parameter & Symbol & Value\\
\hline
Mass of primary & $M_{\rm A}$ & $1.7$M$_{\odot}$ \\
Mass of secondary & $M_{\rm B}$ & $0.2$M$_{\odot}$ \\
Semi-major axis & $a$ & 207~au \\
Eccentricity & $e_\mathrm{b}$ & 0.32 \\
Inclination & $i_{\rm b}$ & 49$^{\circ}$ \\
Position angle (of ascending node) & $\Omega$ & 47$^{\circ}$ \\
Argument of periapsis & $\omega$ & 18$^{\circ}$ \\
\hline
Outer radius of gas disc & $R_{\rm out}$ & 60~au \\
Disc mass & $M_{\rm d}$ & 0.003~M$_{\odot}$ \\
Position angle (of ascending node) & $\Omega_{\rm d}$ & 183.5$^{\circ}$ \\
Inclination & $i_{\rm d}$ & 15.9$^{\circ}$ \\
Aspect ratio & $H/R$ & 0.05 \\
Viscosity & $\alpha_{\rm SS}$ & 0.005 \\
\hline
\end{tabular}
\end{table}

\begin{figure*}
    \centering
    \includegraphics[width=\textwidth]{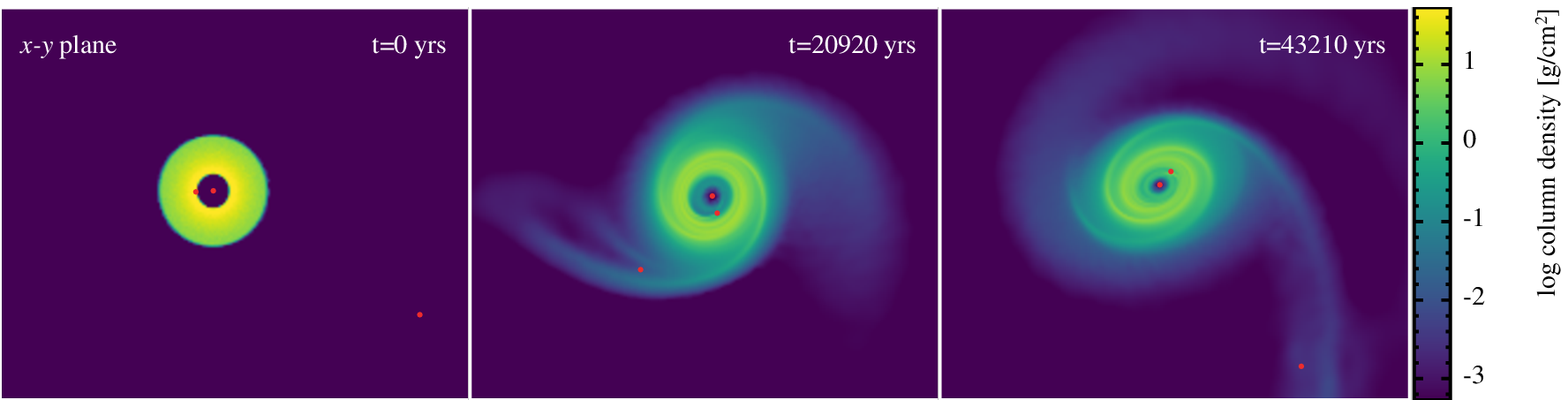}
    \includegraphics[width=\textwidth]{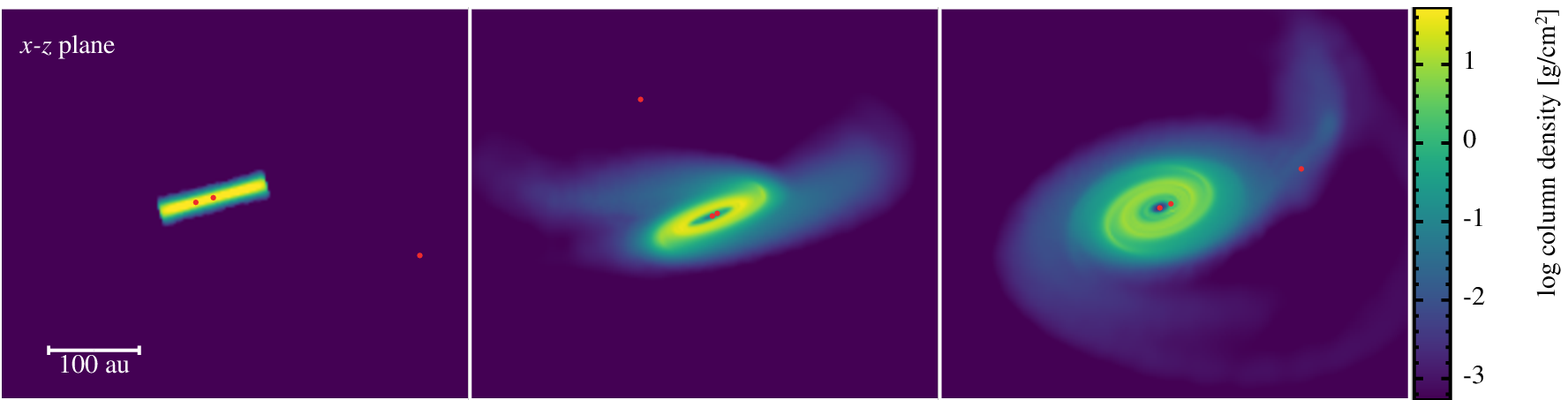}
    \caption{Column density rendering of our simulation with a $5$~M$_{\rm J}$ companion located at $20$~au at $t \approx 0$ (left), $10$ (middle) and $20$ (right) outer binary orbits in the \emph{x-y} (upper) and \emph{x-z} (lower) planes. The stars and planet are indicated with red circles and the plane of the sky corresponds to the \emph{x-y} plane. The accretion radius in our simulation is $5$~au and we do not include an inner disc in our initial conditions.}
    \label{fig:render}
\end{figure*}

%-------------------------------------------------------------
\subsection{Properties of the inner companion}
\label{section:planet_properties}

In order to be consistent with the inner edge location, \citet{vanderPlas:2019gy} suggest that an inner companion would have a mass of $0.01$-$0.1\,$M$_{\odot}$ and must be located around $13$~au. Additionally, the mass of this companion must be low enough that it does not leave a kinematic signature that reveals its presence \citep[as in][]{Pinte:2019xo,Pinte:2020gh}.  On this basis, \citet{Facchini:2017of} suggest the inner companion be less massive than Jupiter, or else it will produce a detectable gap in the $^{12}$CO observations in Figure 4 of \citet{Rosotti:2020nj}. However, this estimate does not take into account any potential inclination of the companion's orbit, the relatively small orbit of the inner companion or the phase of its orbit --- all of which can affect the kinematic signature left by the planet.

We thus conducted numerical simulations to estimate the properties of the inner companion using the smoothed particle hydrodynamics code \textsc{Phantom} \citep{Phantom}. For the outer companion we adopted the orbital parameters from the best fit of \citetalias{Gonzalez:sub} as listed in Table~\ref{table:properties}, while for the inner companion we considered $M_{\rm C} = 5$, $10$ or $20$~M$_{\rm J}$ located at $R_{\rm C} = 10$, $15$ or $20$~au (these masses represent a compromise between the higher mass predictions from \citealt{vanderPlas:2019gy} and lower mass from \citealt{Facchini:2017of}). We examined both how the inner edge of the outer disc evolved and the $^{12}$CO J=3-2 channel maps after 10 orbits of the outer binary (corresponding to $2.2 \times 10^4$~yrs). We compare our results to the observed location of the inner edge of the outer disc \citep{Wagner:2015ph,Long:2018vt,vanderPlas:2019gy} and the CO $J=3-2$ channel maps presented in Figure~9 of \citetalias{Gonzalez:sub}.

\subsubsection{Hydrodynamics}
\label{sec:inneredge}
We adopt the numerical parameters of the best fit orbit in \citetalias{Gonzalez:sub} with several alterations. We decrease the accretion radius of both stars to be 5~au and the disc is initially set up between 20~au and 60~au. The outer radius in our simulation is initially set to be larger than the outer radius observed in mm dust \citep[$\sim 40$~au,][]{Long:2018vt,vanderPlas:2019gy}, allowing the outer edge to be naturally truncated by tidal interaction with the outer binary as well as radial drift. We do not explicitly model the inner disc as this is quite computationally expensive, but do not prevent gas from moving interior to the planet orbit. The planet is started on a circular orbit with an orbit in the plane of the disc with an accretion radius of $0.25$~$R_{\rm Hill}$ \citep{Nealon:2018ic}. As in \citetalias{Gonzalez:sub}, the disc is modelled with $N=10^6$ particles (corresponding to $\sim2.5$ smoothing lengths per scale-height). As we are only interested in the signature that may be present due to the inner companion, we do not consider dust in these hydrodynamical simulations and simply assume for the radiative transfer that the dust and gas are well coupled in making these maps\footnote{We refer to \citetalias{Gonzalez:sub} for a more thorough investigation of the kinematics, including the effect of multiple dust grains.}.

\begin{figure*}
    \centering
    \includegraphics[width=\textwidth]{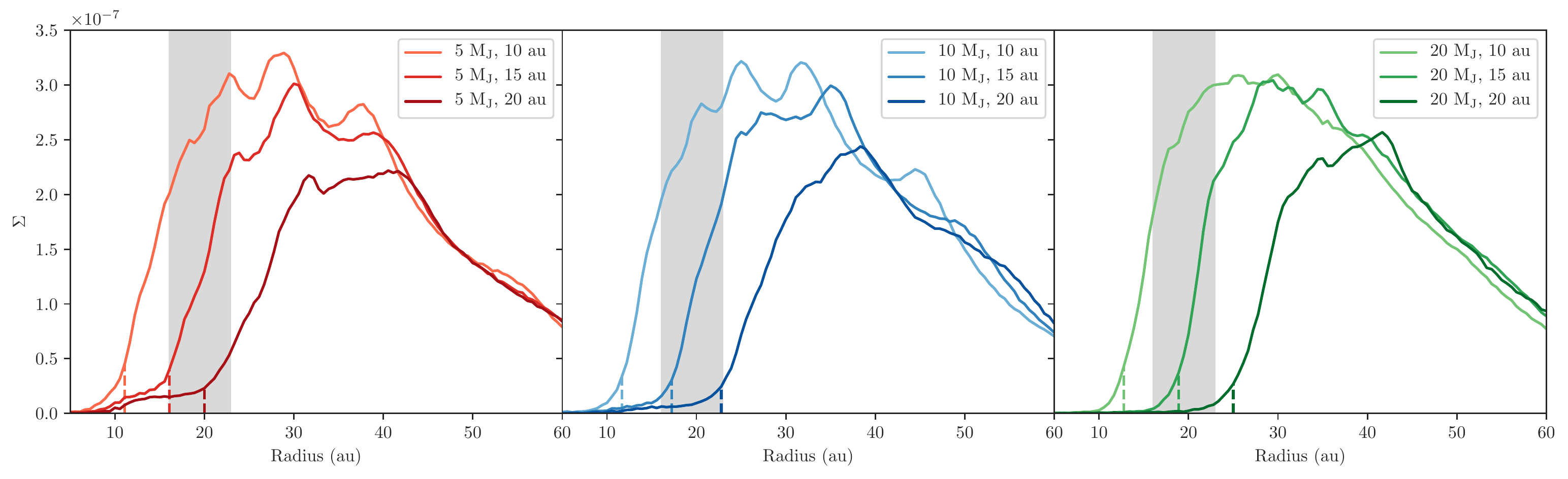}
    \caption{Surface density profiles from our simulations after $2.2\times 10^4$~yrs (10 outer binary orbits or $>300$ inner companion orbits). We consider companion masses of 5, 10 and 20M$_{\rm J}$ located at 10, 15 and 20~au. We measure the inner edge to be where the surface density drops below $10\%$ of its maximum value, shown with the small vertical dashed lines near the bottom of each panel. The grey region shows the location of the inner radius from a combination of mm observations \citep[23~au,][]{vanderPlas:2019gy}, scattered light \citep[21~au,][]{Wagner:2015ph} and SED fitting \citep[$18\pm2$~au,][]{Long:2018vt}. For a disc viscosity of $\alpha=5\times 10^{-3}$, for the lower companion masses of $5-10$M$_{\rm J}$ the companion must be located around 15-20~au, for the larger $20$M$_{\rm J}$ it may be located at 15~au.}
    \label{fig:C_OD_P_sigma}
\end{figure*}

The simulations are evolved for 20 orbits of the outer binary (twice as long as in \citetalias{Gonzalez:sub}), corresponding to more than 600 orbits for the inner companion. We present surface density profiles and kinematics at 10 binary orbits ($2.2 \times 10^4$~yrs) in Figure~\ref{fig:C_OD_P_sigma} to be consistent with \citetalias{Gonzalez:sub}. Figure~\ref{fig:render} shows the column density of the simulation with $5$~M$_{\rm J}$ at $20$~au at 0, 10 and 20 binary orbits for comparison.

We measure the inner edge of the outer disc (shown in Figure~\ref{fig:C_OD_P_sigma} with vertical, dashed lines) where the surface density profile drops below $10\%$ of the maximum surface density. We find five cases that show agreement between the inner edge of the disc in our simulations and the range from observations. Our results suggest a consistent inner edge location for the $5$ and $10$~M$_{\rm J}$ located at $15-20$~au or $20$~M$_{\rm J}$ at $15$~au. This confirms the location predicted by \citet{vanderPlas:2019gy} but for lower masses than they postulated.

As gap opening is known to be easier at lower viscosities \citep{Duffel:2013mu}, the depth and width of the gap in our simulations is dependent on the viscosity. For these simulations we have adopted $\alpha=5 \times 10^{-3}$ as in \citetalias{Gonzalez:sub}. However the viscosity in protoplanetary discs may be as low as $\alpha \sim 10^{-4}$ \citep[e.g.][]{bai_stone_2012,Flaherty:2017do,Teague:2018vw}. Analytical approaches considering gap opening by a planet have shown that the minimum mass required to open a gap in the disc is $\propto \sqrt{\alpha}$ \citep[e.g.][]{Crida:2006bv,Dipierro:2017kq}. Thus for a lower viscosity of, say, $\alpha \sim 10^{-4}$, our results would suggest a minimum mass of $M_{\rm C} \gtrsim 0.71~M_{\rm J}$ to truncate the inner disc at the observed radius \citep[18-23~au,][]{Wagner:2015ph,Long:2018vt,vanderPlas:2019gy}. While our conclusions of the planet mass depend strongly on the $\alpha$ in the disc, the results from our timescale comparison in Section~\ref{section:analytical} remain unaffected for different $\alpha$ as long as the disc remains wave-like (with $\alpha \lesssim H/R$) --- as is expected for protoplanetary discs \citep[e.g.][]{Flaherty:2015is,Flaherty:2017do,Pinte:2016oa}.

%\rn{note: any companion of more than $5.2$ jupiter mass at 10au will have more AM than the outer disc}

\begin{figure}
    \centering
    \includegraphics[width=\columnwidth]{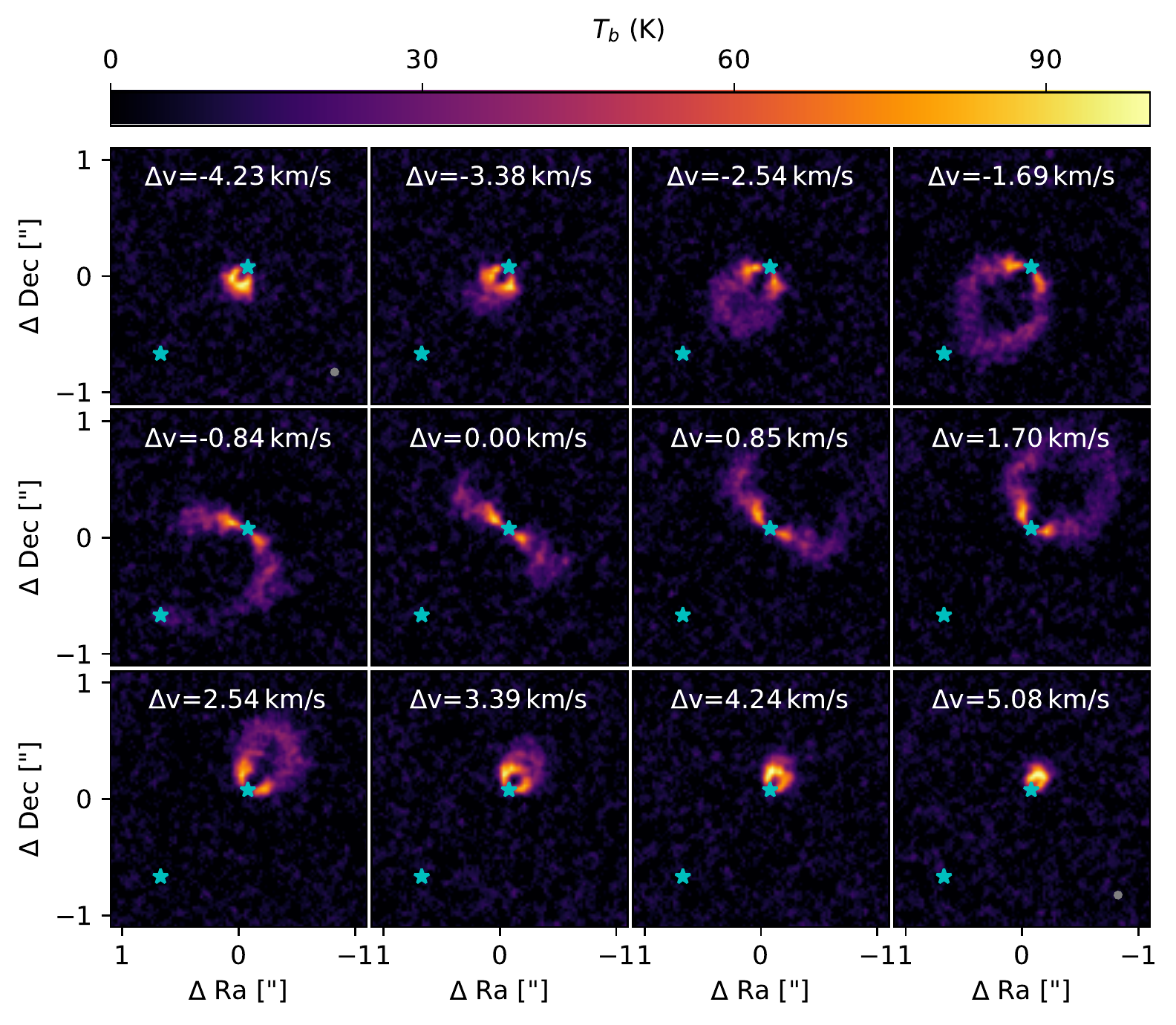}
    \includegraphics[width=\columnwidth]{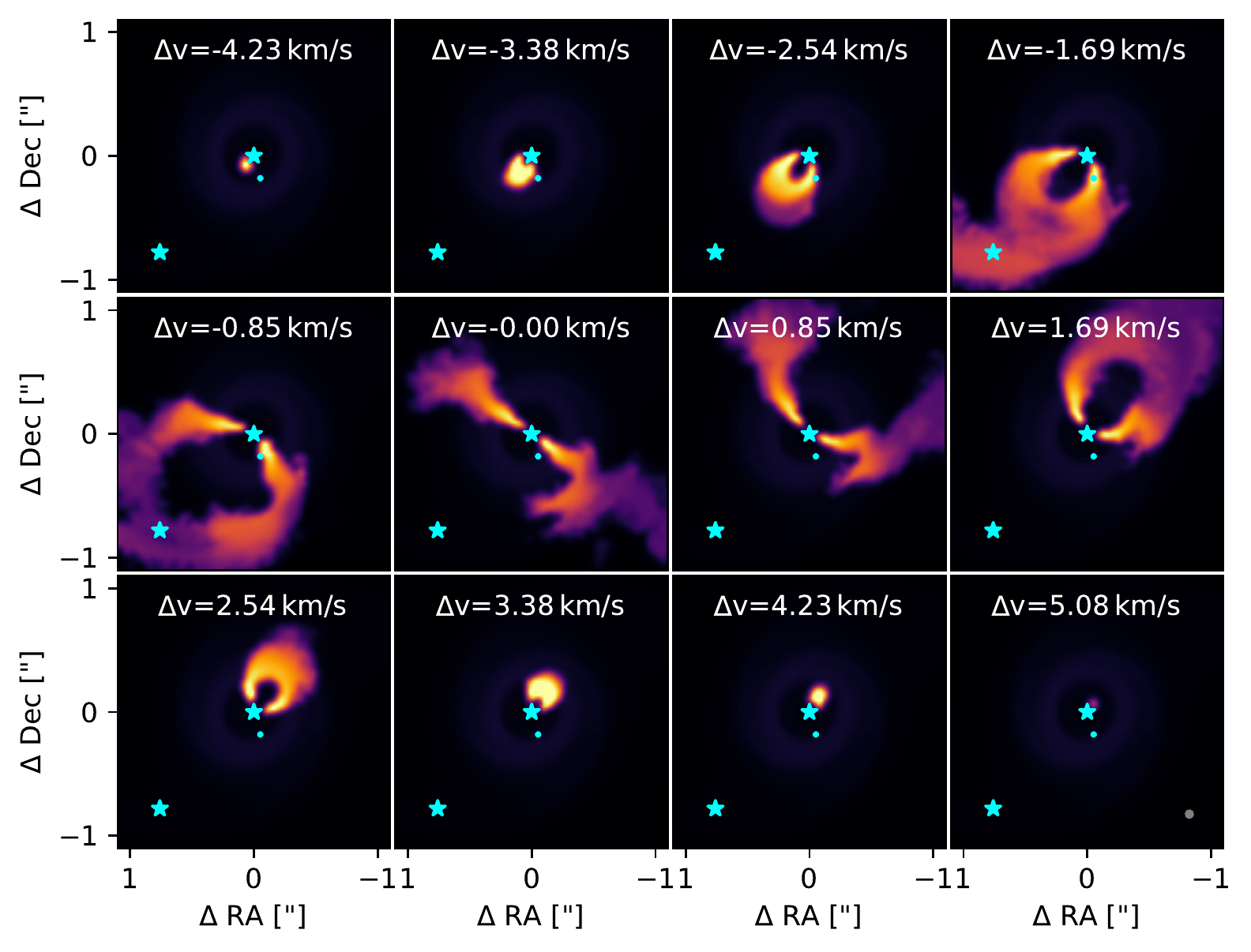}
    \includegraphics[width=\columnwidth]{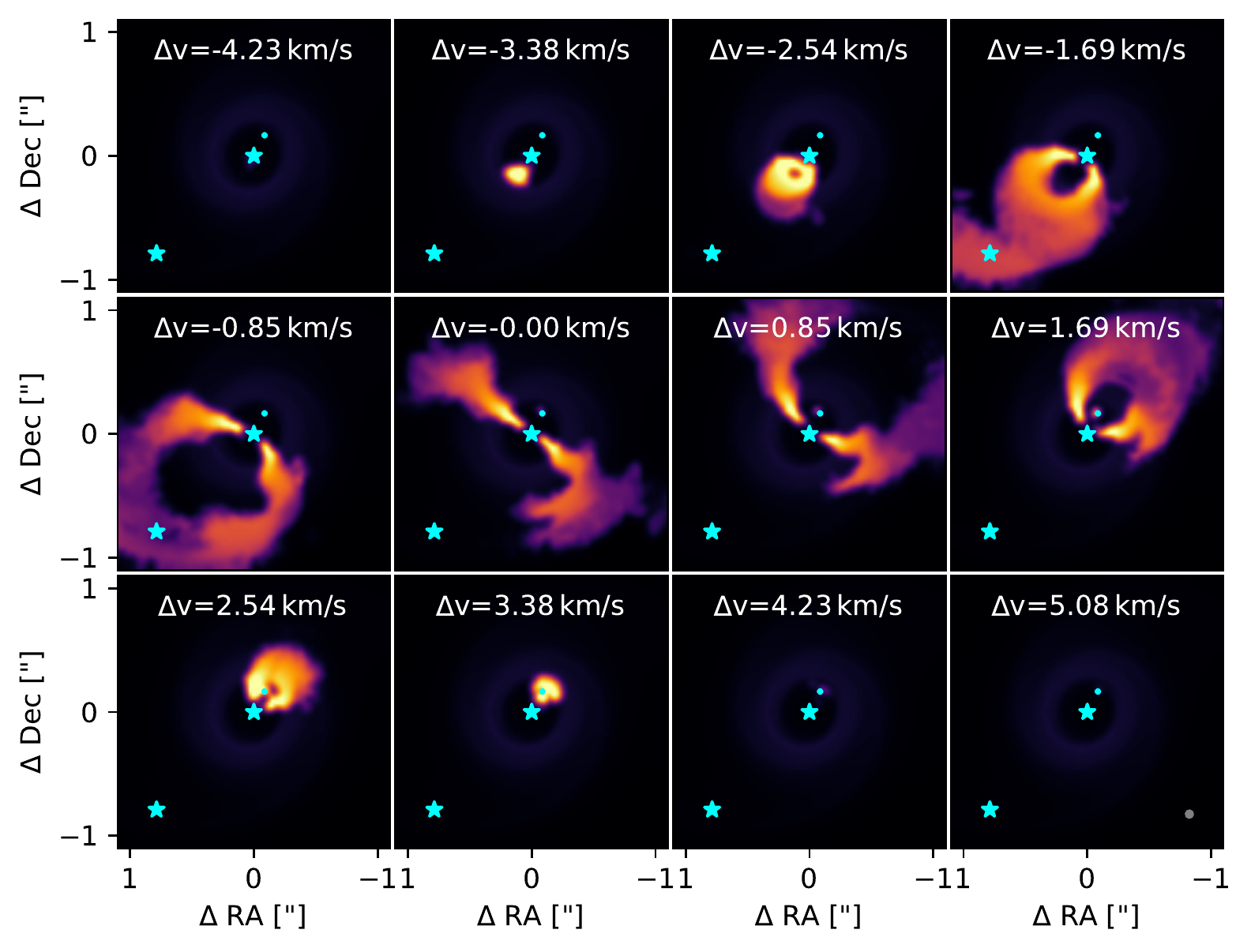}
    \caption{Channel maps of the $^{12}$CO $J = 3-2$ line from ALMA observations (upper), our simulations with $5\,M_{\rm J}$ (middle) and $10\,M_{\rm J}$ (lower) at $20$~au. The stars and planet are indicated in cyan, the beam used for convolution in the bottom right panel in grey and velocities quoted relative to the systemic velocity of $5.12$km/s. All panels are shown with the same spatial and temperature scale. The kinks from either planet are detectable but difficult to distinguish from the spirals driven by the outer binary companion.}
    \label{fig:CO_map}
\end{figure}

\subsubsection{Kinematics}
We use \textsc{mcfost} \citep{Pinte:2006nw,Pinte:2009ye} to calculate the $^{12}$CO $J=3-2$ molecular line emission for each of the simulations in Figure~\ref{fig:C_OD_P_sigma}. Here we use a Voronoi tessellation built around the SPH particles with $10^8$ photon packets. The dust grains assumed by \textsc{mcfost} to be spherical and homogeneous (according to Mie theory), in thermal equilibrium and with dust opacities that are independent of temperature. The grains are distributed across 100 sizes between 0.03 and 1000$\mu$m with a power-law exponent of -3.5. We also assume a uniform CO-to-H2 ratio of $10^{-4}$ for the moment maps. Consistent with \citetalias{Gonzalez:sub}, we take into account CO freeze-out when $T<20$K as well as photo-dissociation and photo-desorption when the ultraviolet radiation is large \citep[Appendix B,][]{Pinte:2018ah}. For the primary star we set $T_{\rm A} = 7250$K with $L_{\rm A} = 6.2$L$_{\odot}$ and for the secondary $T_{\rm B} = 3250$K with $L_{\rm A} = 0.06$L$_{\odot}$. The channel maps are produced with $0.042$km/s resolution, Hanning smoothed consistent with the observed spectral resolution and are convolved with the ALMA CLEAN beam of $0.054 \times 0.052$ mas \citep[e.g.][]{Rosotti:2020nj}.

Figure~\ref{fig:CO_map} shows the channel maps for our two simulations with 5 and $10$~M$_{\rm J}$ at 20~au (we do not include the $20$~M$_{\rm J}$ as it is similar to the $10$~M$_{\rm J}$). For reference, in the upper panel we have reproduced the ALMA channel maps and refer to \citet{Rosotti:2020nj} for details of its calibration. As in \citetalias{Gonzalez:sub}, the spiral arms driven by the outer binary produce features that are particularly noticeable in the lower velocity channels. The difference in the temperature scale between the ALMA channel maps and the lower panels is most likely due to the innermost disc; in the observations the inner disc is able to intercept some flux but in our simulations this is poorly resolved and so the outer disc is brighter than expected. We note that this does not affect either the shape or location of the structures identified, so does not alter our conclusions (see also \citetalias{Gonzalez:sub}). While the deviations due to the planets (indicated in cyan) are identifiable if the planet location is known, they are difficult to distinguish on the background of structure generated by the spiral arms. For the same position of the binary orbit the planets are also located in different azimuthal positions and this will affect the kink signature produced.

For the $10$~M$_{\rm J}$ case there is lower emission in the high velocity channels ($-4.23,4.23$ and $5.08$~km/s). These high velocities correspond to the region close to the accretion radius of the primary star set in our simulations. In the $10$~M$_{\rm J}$ the higher mass planet accretes more gas than the lower one, preventing the build up of a significant inner disc and hence emission associated with this high velocity gas. As the channel maps shown in \citetalias{Gonzalez:sub} presented from ALMA show emission at these velocities which is more consistent with the $5$~M$_{\rm J}$ mass planet, we favour the lower planet mass of $5$~M$_{\rm J}$ for the inner companion. Comparison between the 5~M$_{\rm J}$ and 10~M$_{\rm J}$ cases suggests that there will be even more emission in the higher velocity channels for a lower mass planet, which would be more similar to the ALMA observations. We thus place an upper limit on the planet mass of $5$~M$_{\rm J}$ but suggest that is it likely to be lower than this.

\section{Long term evolution of HD~100453}
\label{section:long_term_evolution}

Figure~\ref{fig:HD100-sketch} shows the full picture of HD~100453 including the inner disc, inner companion, outer disc and outer companion --- each misaligned to the other relative components. With these in mind we consider the long term evolution of HD~100453 and focus on the relative misalignment between the inner and outer disc. As the inclination damping time-scale for a planet of a few Jupiter masses is quite rapid \citep{Xiang-Gruess:2013fg,Bitsch:2013hg}, it is not feasible that the $72 \degr$ relative misalignment between the inner and outer disc is caused only by the inner companion. We thus seek to explain the misalignments in HD~100453 using the outer, bound companion. Section~\ref{section:planet_properties} suggests that the planet has a mass of $\lesssim 5$~M$_{\rm J}$. Here we note that although a lower planet mass could be successfully hidden in the channel maps of Figure~\ref{fig:CO_map}, it would require a lower viscosity than we have used in our simulations. To be consistent with our simulations we thus adopt the lowest mass used there of $5$~M$_{\rm J}$ located at $20$~au, but note that the planet mass could be lower than this.

\subsection{Kozai-Lidov oscillations}
\label{subsection:nbody}
The $\sim 61^{\circ}$ relative misalignment between the outer disc and binary plane clearly meets the criteria for the Kozai-Lidov mechanism \citep{Kozai:1962gq,Lidov:1962gq}. Kozai-Lidov oscillations occur for small bodies inclined by more than $39.2\degr$ to an external companion, where conservation of the angular momentum perpendicular to the binary orbit causes an exchange between eccentricity and inclination in the small body. For a rigid disc, this phenomenon occurs on a time-scale of \citep{Martin:2014bb}
\begin{align}
    \langle T_{\rm KL} \rangle \approx \frac{(4 - p)}{(5/2 - p)} \frac{ \sqrt{M_{\rm A} M}}{M_{\rm B}} \left(\frac{a}{R_{\rm out}} \right)^{3/2} {T_{\rm b}} \,\,\,,
    \label{equation:average_KL}
\end{align}
where $p$ is the index of the surface density profile power law and $M = M_{\rm A} + M_{\rm B}$. As noted by \citet{Martin:2014bb}, Equation~\ref{equation:average_KL} does not take into account any inclination dependence and so is only accurate up to a factor of a few. The Kozai-Lidov oscillation period additionally depends on the aspect ratio, viscosity and binary eccentricity \citep{Fu:2015wg,Franchini:2019gc}. Despite this we can use Equation~\ref{equation:average_KL} to estimate whether Kozai-Lidov oscillations are relevant to the evolution of HD~100453. Assuming an unbroken disc and using the values in Table~\ref{table:properties} and $p=1$, $\langle T_{\rm KL} \rangle \approx 1.1 \times 10^5$~yrs. If the disc was continuous and unbroken, the entire disc of HD~100453 would oscillate every $1.1 \times 10^5$~yrs. However, the observations of HD~100453 clearly show the inner and outer disc are disconnected. Due to the strong radial dependence on the torque exerted by the outer companion the disconnected discs and inner companion will oscillate differentially, naturally leading to a range of relative misalignments.

We use the $N$-body code \textsc{Rebound} \citep{REBOUND, Rein:2015nj} to show how the Kozai-Lidov mechanism can generate such misalignments on long timescales. Here we assume that both discs can be modelled as a rigid body (justified by their limited radial extent, i.e. from 0.1-1~au and 21-40~au), allowing each to be modelled by a test particle. Each particle is located at the radius where it has the same Kozai-Lidov frequency as the radially extended disc would have. From \citet{Martin:2014bb}, this corresponds to a semi-major axis of
\begin{align}
    a_{\rm p} = \left( \frac{5/2 - p}{4-p} \right)^{2/3} \left(1-e_{\rm b}^2 \right) R_{\rm out}\,\,\,,
    \label{equation:semimajoraxis}
\end{align}
where $e_{\rm b}=0.32$ (Table~\ref{table:properties}). This corresponds to $a_{\rm p}=0.56,22.6$~au for the inner and outer disc respectively. We caution that the above approximation does not take into account the location of the inner edge of the disc. For simplicity we initialise each particle assuming it is in the plane of the outer disc (using the disc position and inclination angle from Table~\ref{table:properties}), but note that the inner disc is more strongly misaligned to the binary than this.

Figure~\ref{fig:nbody} shows the evolution of the misalignment of the inner disc, inner companion and outer disc for the lifetime of HD~100453. Due to the differential torque applied by the Kozai-Lidov mechanism relative misalignments of less than 20$\degr$ between both the inner and outer disc as well as the inner companion and inner disc naturally occur. In this representation the inner disc does not appear to evolve, but that is because the Kozai-Lidov time-scale ($2.8 \times 10^8$~yrs) is much longer than the age of the system. The outer disc has the most rapid evolution with oscillations every $\sim 2.5\times10^5$~yrs and the inner companion oscillates only slightly slower than this. This difference is only due to the different distances between the outer companion and each component, essentially causing the outer disc and inner companion to oscillate around the practically stationary inner disc. The growth in the magnitude of the oscillations in tilt in the outer disc are due to the eccentricity of the outer companion \citep{Li:2014ua}. At all times during the evolution, the relative misalignment between the binary and each component remains above the Kozai-Lidov threshold of $39.2 \degr$.

We caution that an N-body approximation neglects viscous and pressure effects of the gas and this will alter the evolution of the relative misalignment over time \citep{Martin:2014bb,Dipierro:2018qp}. Hydrodynamic simulations by \citet{Martin:2014bb} showed that dissipation within the fluid disc causes the oscillations to damp, so it is not likely that the outer disc in HD~100453 will undergo as many oscillations as Figure~\ref{fig:nbody} predicts. As we shall show in Section~\ref{section:rel_disc_misalignment}, in this case differential precession can still cause the outer disc to precess faster than the inner disc and hence the observed misalignment.

\subsection{The inner companion}
\begin{figure*}
    \centering
    \includegraphics[width=\textwidth]{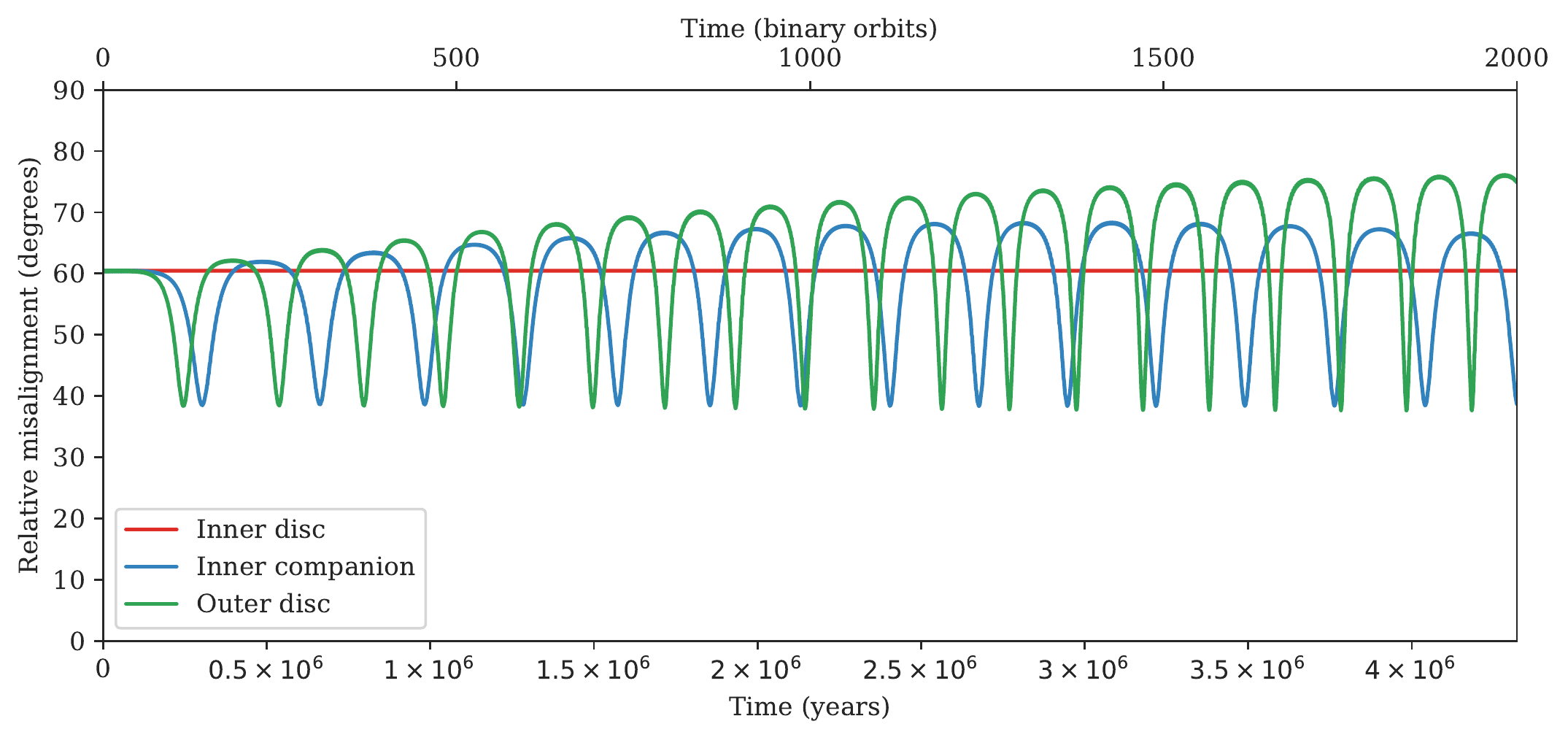}
    \caption{$N$-body representation of the long-term evolution of HD~100453, including an inner disc, outer disc and inner companion. The misalignment here is measured with respect to the binary orbital plane. The Kozai-Lidov mechanism routinely oscillates the components around the primary, driving a relative misalignment of $-20$ to $+15\degr$ between the inner and outer disc.}
    \label{fig:nbody}
\end{figure*}

\begin{figure}
    \centering
    \includegraphics[width=\columnwidth]{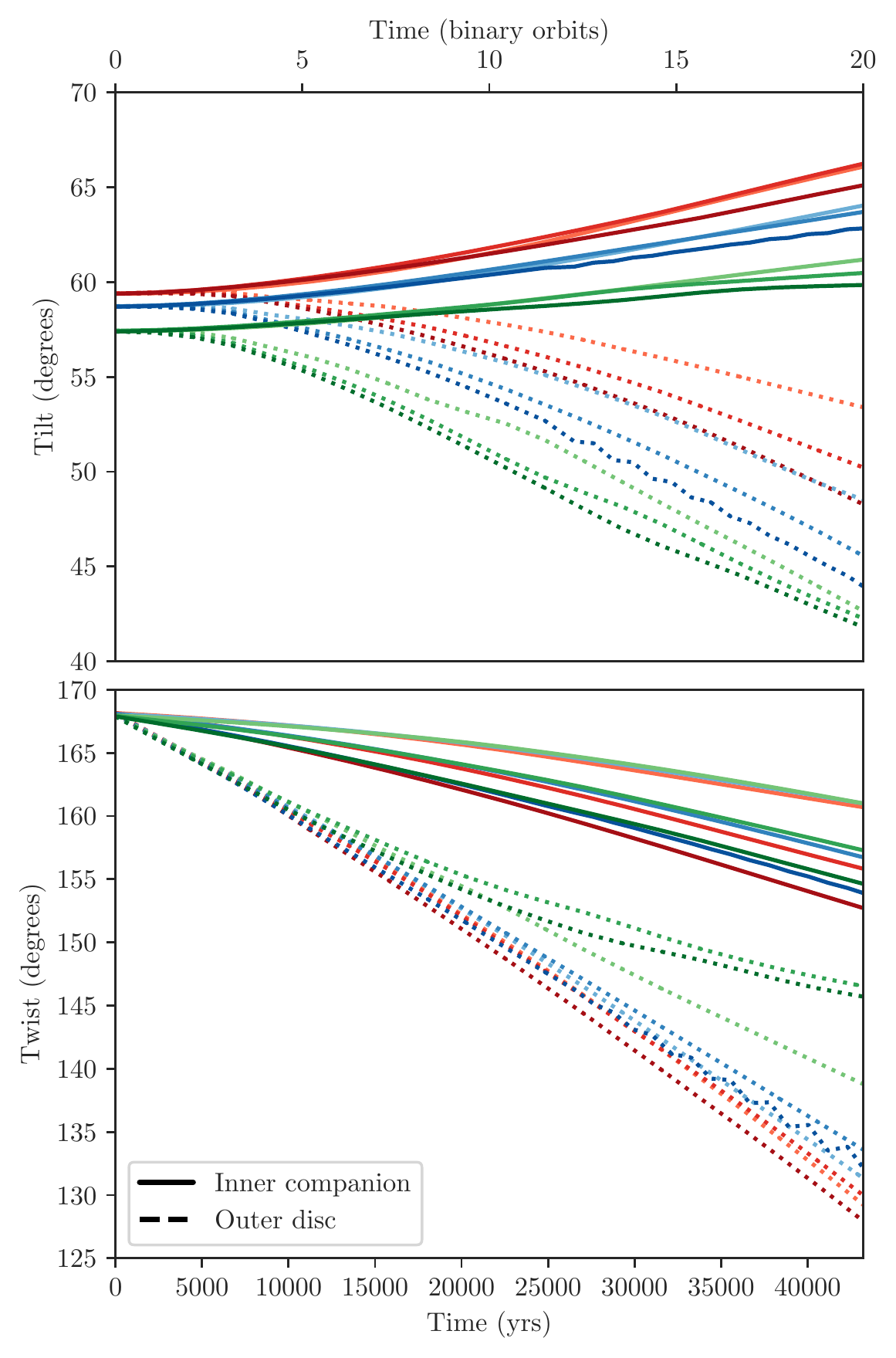}
    \caption{Tilt $\beta$ (upper) and twist $\gamma$ (lower) of the inner companion and outer disc across our hydrodynamic simulations, with the same colour scheme as in Figure~\ref{fig:C_OD_P_sigma}. The different tilt and twist rate indicate that the planet decouples from the orbital plane of the outer disc. As the tilt and twist are measured from the total angular momentum, the initial tilt is different for each of the planet masses.}
    \label{fig:tilt_and_twist}
\end{figure}

For the outer disc mass observed in HD~100453 and the masses of the inner companion used in Section~\ref{section:planet_properties}, \citet{Martin:2016qf} predicts that on long time-scales the inner companion will circulate with a precession rate and tilt that is independent of the outer disc. In Figure~\ref{fig:tilt_and_twist} we show the evolution of the tilt and twist of the inner companion and the outer disc across $4.3 \times 10^4$~yrs (20 outer binary orbits). The tilt $\beta(t)$ and twist $\gamma(t)$ are calculated from the components of the unit angular momentum $\Bell(t)$ as
\begin{align}
    \beta(t) = \cos^{-1} (\Bell_z(t))\,\,\,,
    \gamma(t) = \tan^{-1} \left( \frac{\Bell_y(t)}{\Bell_x(t)} \right)\,\,\,,
\end{align}
where we have rotated the simulation so that $\Bell_{z}$ is parallel to the total angular momentum vector (and not the plane of the sky) and we use a weighted average to calculate the unit angular momentum vector of the outer disc. As predicted, Figure~\ref{fig:tilt_and_twist} shows that while the relative misalignment of the outer disc decreases towards the binary plane, the planet increases its relative misalignment, moving away from the plane of the outer disc. This is in agreement with behaviour found in \citet{Martin:2016qf} and \citet{Franchini:2020ht}. Although such behaviour was not found in our N-body calculation (see Figure~\ref{fig:nbody}), this is likely due to the different masses used, which affect the time-scale of the resulting oscillations. The precession rate of the inner companion is also slower than that of the outer disc. We additionally find that the eccentricity of the planet increases over the course of the simulations, reaching a maximum (for the 20M$_{\rm J}$ located at 20~au) of $e \sim 0.03$ by the end of the simulation. Towards the end of the simulations the eccentricity decreases for the lowest planet mass cases.

Despite decoupling from the outer disc, the inner companion is still able to dynamically set the inner edge of the disc through dynamic friction \citep{Rein:2012so}. This behaviour has been observed for inner companions with a mass as low as $\sim$Jupiter mass \citep{Picogna:2015bo} and so is expected to be consistent for the lower mass estimates given in Sect.~\ref{sec:inneredge}. On longer time-scales, \citet{Martin:2016qf} also predicts that the inner companion will undergo Kozai-Lidov oscillations but they will not be damped as is the case for the outer disc. This may increase the relative inclination between the planet and disc enough that the gap is unable to be maintained \citep{Martin:2016qf}. Additionally, the planet orbit may even become retrograde \citep{Li:2014ua,Franchini:2020ht}.

%-------------------------------------------------------------
%-------------------------------------------------------------
\subsection{Precession}
\label{section:rel_disc_misalignment}
In addition to regular oscillations by the Kozai-Lidov mechanism driven by the outer binary, the entire HD~100453 system is continuously precessing around the orbit of the binary. This precession occurs due to the non-Keplerian terms in the potential in the presence of the outer companion, irrespective of whether the Kozai-Lidov mechanism is acting or not. In the frame of the primary, this results in differential precession of the inner disc, inner companion and outer disc. This precession alters the twist of the inner and outer disc and will thus alter the relative misalignment that occurs between the two. Here we consider the rate of precession of each component of HD~100453 in turn and compare this to the respective Kozai-Lidov timescales.

We start with the precession rate of the outer disc: using Eq.~\ref{equation:tp} and assuming a misalignment between the outer disc and outer companion of $61\degr$ yields a precession rate of $8.8 \times 10^5$~yrs. For the outer disc, $\langle T_{\rm KL} \rangle \lesssim T_{\rm p}$ and the precession timescale is slightly longer than the Kozai-Lidov timescale. This suggests that the observed orientation of the outer disc has been driven by both differential precession and the Kozai-Lidov mechanism.

Next we consider the precession rate of the inner companion. Our hydrodynamical simulations confirmed previous predictions that the inner companion and outer disc will decouple and precess independently. Assuming the inner companion was initially co-planar with the disc (as in our simulations), the precession rate of the inner companion can be expressed relative to the precession rate of the outer disc as \citep{Picogna:2015bo}
\begin{align}
    \left( \frac{T_{\rm OD}}{T_{\rm C}}\right)_{t=0} = 
    2 \left(\frac{R_{\rm C}}{R_{\rm out}} \right)^{2/3}\,\,\,,
\end{align}
where $T$ is the period of the precession and $t=0$ indicates that this estimate is based on the initial conditions used in our simulations. The above suggests that the inner companion will precess every $1.1 \times10^6$~yrs, Figure~\ref{fig:tilt_and_twist}. This is about an order of magnitude longer than the Kozai-Lidov time-scale for the inner companion, with $\langle T_{\rm KL} \rangle \lesssim T_{\rm p}$. With the Kozai-Lidov mechanism occurring on a much faster time-scale than precession, it is likely that the plane of the orbit, inclination and eccentricity of the inner companion are the result of Kozai-Lidov oscillations.

Finally we consider the innermost disc. As the inner disc has the smallest angular momentum of the system, it will be driven by both companions and the outer disc. However due to its proximity and mass the inner companion will dominate whenever there is a misalignment between it and the inner disc. With an extent between $0.1-1.0$~au and a relative misalignment of $20\degr$ for simplicity (e.g. Figure~\ref{fig:nbody}), the precession rate is $6\times 10^6$~yrs. Thus for the inner disc $\langle T_{\rm KL} \rangle > T_{\rm p}$, however we note that the precession time-scale for the inner disc is comparable to the age of HD100453~A.

\subsection{Alignment}
\label{section:timescales}
Hydrodynamical effects will seek to align the components of HD~100453 over time. In \citetalias{Gonzalez:sub} we addressed this concern regarding the outer companion and the outer disc, noting that the time-scale for the disc to realign is roughly the viscous time-scale ($T_{\nu} = R^2/\nu$) and thus on the same order as the lifetime of the disc itself.

Such effects will also seek to align the inner disc, inner companion and outer disc over time. The frequent oscillation of the inner companion and outer disc by the misaligned outer companion prevents alignment of these two components. Alignment between the inner companion and the inner disc can be estimated in terms of the precession time-scale from \citep{Bate:2000fk}
\begin{align}
\frac{T_{\rm align}}{T_{\rm p}}\sim\frac{1}{K\cos {\Delta i}\,q\,\alpha_{\rm SS}} \left(\frac{H}{R}\right)^2\left(\frac{R_{\rm out}}{a}\right)^{-3}\,\,\,,
\end{align}
where $q$ is the mass ratio between the inner companion and the primary star, $R_{\rm out}$ is the outer edge of the inner disc, ${\Delta i}$ is the relative misalignment between the inner disc and inner companion and $K$ is derived in Equation~\ref{Eq:K_Terquem}. Using the estimates for the disc size from above, we find that the alignment time-scale of this disc is longer than the precession time-scale. However the inner disc is certain to accrete on a time-scale faster than this; assuming $\alpha_{\rm SS} = 5 \times 10^{-3}$ the viscous time-scale corresponds to $9.8 \times 10^3$~yrs. The viscosity is likely to be lower than this, which would increase the viscous time of the inner disc. For an $\alpha_{\rm SS} = 5 \times 10^{-4}$, the viscous time would increase to $9.8 \times 10^4$~yrs. This is the most rapid of all the time-scales considered so far and strongly suggests that the inner disc is being fed slowly from the outer disc (as in our Figure~\ref{fig:C_OD_P_sigma} for the lower planet mass).

%\ncc{Silly speculation: I know that you need to disconnect the ID and the OD (ang. mom.-wise) so this mechanism can work. This is nicely done by the IC and seems to stop material inflow. But, could you have an ID that after disc breaking/tearing is progressively becoming less extended until it reaches a steady state/size given by the equilibrium between accretion and replenishment from the OD? It seems unlikely that NO material at all falls into the cavity from the OD. Given the BUM BUM thing and the highly misaligned geometry of all the components some material should be able to fall. Perhaps not once per orbit, but from time to time, the inner regions get replenished thanks to the BUM BUM effect. I do not know if this can be checked but we could at least a bit speculate about it.}

%-------------------------------------------------------------
%-------------------------------------------------------------

\section{Discussion}
\label{section:disc}
%-------------------------------------------------------------

\subsection{The complete picture}
We have argued that HD~100453 must have a hidden, inner companion. While such a companion was suspected in observations \citep{Wagner:2015ph,vanderPlas:2019gy,Rosotti:2020nj}, our argument is from the dynamics of the broken disc. This additional component self consistently explains the origin of the misalignments in HD~100453 using the misaligned outer companion. HD~100453~B drives the outer disc, planet and inner disc to precess and occasionally undergo Kozai-Lidov oscillations. By itself, the Kozai-Lidov mechanism is not able to drive the required $72\degr$ misalignment between the inner and outer disc --- it is predominantly differential precession that results in such a strong misalignment. Assuming the current radial extent of the discs, we found that for both the outer disc and inner companion the Kozai-Lidov time-scale was shorter or comparable to that for differential precession. For the inner disc both the precession and Kozai-Lidov timescales were comparable to the age of the system but its accretion time-scale is the most rapid of the whole system.

Assuming that the inner disc, inner companion and outer disc are originally co-planar, this suggests the following chain of events for HD~100453:

\begin{enumerate}
    \item After formation of the inner companion, a combination of differential precession and the Kozai-Lidov mechanism drives the inner disc, inner companion and outer disc to precess. The outer disc precesses most rapidly and the inner disc is not meaningfully perturbed, causing a relative misalignment to develop between the two.
    \item The inner companion is decoupled from the outer disc, with an independent precession rate and orbital plane.
    \item Subsequent Kozai-Lidov driven oscillations and differential precession continued the evolution of the outer disc's orientation, enhancing the relative misalignment between the inner and outer disc.
\end{enumerate}

When acting on a fluid disc the Kozai-Lidov mechanism is predicted to damp after several oscillations, leaving a disc with no noticeable eccentricity and a large relative misalignment to the outer companion \citep{Martin:2016qf}. On longer time-scales this relative misalignment damps to the critical Kozai-Lidov angle of $39.1 \degr$. This prediction is consistent with the picture we present here; the Kozai-Lidov mechanism damps after several oscillations, leaving behind circular discs that are misaligned. The inner disc is more strongly misaligned to the outer companion than the outer disc \citep{Benisty:2017kq}, suggesting that the outer disc started with a larger relative misalignment to the outer companion and is slowly damping to the critical Kozai-Lidov angle. This misalignment is maintained over a significant fraction of the disc expected lifetime.

This scenario explains the relative misalignment between the inner and outer disc in HD~100453 as a result of the \emph{outer disc precessing faster than the inner disc}. We refer to this as `precession inception' because the outer components are precessing the most rapidly. To our knowledge HD~100453 appears to be unique amongst protoplanetary discs in this behaviour. We refer to discs like HD~142527 and J1604 for contrast; for the former, numerical modelling has shown that an eccentric, misaligned inner stellar companion forms a misaligned inner disc that casts the observed shadows \citep{Price:2018pf}. For the latter, although no companion has yet been found \citep{Sicilia-Aguilar:2020xx} there is evidence that the narrow lane shadows in J1604 are moving, implying precession of the inner disc that is casting them \citep{Pinilla:2018gb}. While the shadows cast by the inner disc in each of these three cases have identical observational characteristics, in HD~100453 the origin of the misalignment (and hence the shadows) is due to the outer disc orientation changing rather than the inner disc.

Importantly, this picture fundamentally only relies on the strong misalignment of the outer companion and that the inner and outer disc are suitably disconnected (and does not require any initial misalignment between the discs). While such misalignments may appear exotic, a misaligned disc is the self-consistent and natural outcome of a system with a misaligned outer companion. Because of differential precession, different configurations to the one proposed for HD~100453 will still result in a relative disc misalignment even if there is no Kozai-Lidov mechanism acting. Strictly then the outcome of a relative disc misalignment only requires the misaligned outer companion and suitably disconnected discs. Alternatively, if the outer companion is not strongly misaligned \citep[as in][]{Wagner:2018hs} then neither precession nor Kozai-Lidov oscillations can be used to explain the relative misalignment. In this case, secular resonances between the inner disc and inner companion can result in a large relative misalignment, but we note that this study neglects damping effects that may hamper this \citep{Owen:2017oj}.

\subsection{Limitations}
\label{section:limitations}
The estimate for the Kozai-Lidov time-scale in a rigid, extended disc in Equation~\ref{equation:average_KL} predicts that $\langle T_{\rm KL} \rangle \sim 1.1 \times 10^5$~yrs. However, our simulation in \citetalias{Gonzalez:sub} that goes for twice this length of time does not show any evidence of these oscillations (either a rapid increase in eccentricity or inclination).

While the Kozai-Lidov mechanism can be suppressed in discs that are massive enough to be self-gravitating \citep{Batygin:2011oj,Batygin:2012ge,Fu:2015wg}, the measured disc mass of HD~100453 precludes this. It is also unlikely to be caused by the numerical method we are using, as \textsc{Phantom} was also used in \citet{Martin:2014bb} and \citet{Picogna:2015bo} which showed the Kozai-Lidov mechanism acting in a fluid disc.

\citet{Lubow:2017hw} found that the aspect ratio of the outer disc can also affect whether Kozai-Lidov oscillations can occur. For a small perturber, they showed that oscillations do not occur when $M_2/M_1  < ((H/R) n/n_{\rm b})^2$, where $H/R$ is measured at the outer edge, $n_{\rm b}$ is the orbital frequency of the binary and $n$ the orbital frequency of the outer disc. For HD~100453 the aspect ratio at the outer disc is $H/R > 0.10$ and thus satisfies this criterion, potentially explaining the lack of oscillations in our simulations.

As noted by \citet{Martin:2014bb}, the estimate used does not take into account the relative inclination between the binary and the disc and so is less likely to be accurate for large misalignments as we are modelling. This estimate additionally does not take into account the location of the inner edge of the disc, the aspect ratio, the disc viscosity or binary eccentricity. Explorations with hydrodynamical simulations by both \citet{Fu:2015wg} and \citet{Franchini:2019gc} have shown that the rate at which the oscillations occur and how quickly they can begin depend sensitively on these parameters. If indeed this estimate is out by a factor of a few, the Kozai-Lidov time-scales will be increased by a factor of a few but the precession estimates (which do take into account inclination) will not change. This will not alter our comparison in Section~\ref{section:rel_disc_misalignment}, as differential precession will still be the fastest time-scale for the inner companion and outer disc. In this instance, conducting simulations over even longer time-scales than we have considered here will resolve this.

We also consider the large spirals formed through the tidal interaction with the outer companion as a potential way to prevent the Kozai-Lidov oscillations from occurring. \citet{Batygin:2011oj} showed that rapid apsidal precession in the disc can prevent the eccentricity growth required for the Kozai-Lidov mechanism to occur. In HD~100453, the strong spiral arms from the outer binary may act in a similar fashion, preventing eccentricity growth. If this is the case, differential precession will still be able to drive the relative disc misalignment (even from a co-planar initial orientation) just on longer time-scales.

The observations of HD~100453 also show no evidence of eccentricity \citep{vanderPlas:2019gy,Rosotti:2020nj}. This is expected if the Kozai-Lidov oscillations have damped, as they are noted to over time for fluid discs \citep{Picogna:2015bo,Martin:2016qf}. Even after damping of the Kozai-Lidov effect, the misalignment of the outer companion will still drive differential precession and (as long as the inner and outer discs are suitably disconnected) drive a relative misalignment between the two discs. This scenario will still result in a strong relative misalignment but no increase of the disc eccentricity.

Our prediction that the outer disc is precessing faster than the inner disc is also based on the current extent of the inner disc. As discussed in Section~\ref{section:recap}, this has been difficult to accurately measure from observations. Specifically, the time-scales derived in Section~\ref{section:long_term_evolution} are quite long because of the separation between the outer edge of the inner disc and the inner companion. Earlier in the evolution of the disc this gap would be narrower, the inner disc would have a larger radial extent and the inferred precession rate would be faster and the accretion time-scale longer. Assuming that the disc and HD~100453~A formed with a similar orientation, the direction of rotation of the primary star could be used to confirm or deny this: if indeed the inner disc has the longer precession time, its orientation (which is known relative to the outer disc) would be closer to the star's rotation axis than the orientation of the outer disc.

%-------------------------------------------------------------
\subsection{Origin of the misaligned outer companion}
\label{section:originofB}

A strongly misaligned binary as in HD~100453 can naturally occur during star formation. Scenarios such as this with a range of misalignments are frequently identified in radiative hydrodynamic simulations of star formation from collapsing molecular clouds \citep{Bate:2018ls,Wurster:2019pb}. Alternatively, the secondary may have been captured and introduced to the system after the formation of the primary. The latter is supported by the difference in ages that has been measured between the primary and the secondary \citep{Collins:2009he,Vioque:2018pg}.

Capture of HD~100453~B is also supported by the development of the misaligned discs. In Section~\ref{section:analytical} we established that the inner and outer disc can only have been separated by the presence of an inner companion. Assuming a core-accretion model of planet formation, \citet{Martin:2016qf} showed that the Kozai-Lidov mechanism prohibits the growth of planetesimals and would thus make it difficult for planets to form in discs undergoing oscillations. This suggests that the planet was likely formed before the outer companion started driving perturbations and thus that it was captured. \citet{Martin:2016qf} alternatively suggests that if the disc is initially more massive such that it is self-gravitating, this will prohibit Kozai-Lidov oscillations until planetesimals form.

%-------------------------------------------------------------
\subsection{Evolution of the narrow lane shadows} \label{sec:shadows}
In the model we present here, the relative misalignment between the inner and outer disc evolves because the outer disc is precessing faster than the inner disc, causing the shadows cast on the outer disc to change, as the orientation of the surface they are being cast on is moving. Previous works have shown that such narrow lane shadows can have important dynamical implications. \citet{Montesinos:2016ig} and \citet{Montesinos:2018zd} have shown that spiral arms can be launched from narrow lane shadows that are stationary or co-rotating with the gas in the outer disc. Although it is remarkable that the spirals appear to be rooted at the locations of the shadows from the inner disc \citep{Benisty:2018ve}, in \citetalias{Gonzalez:sub} we established that these spiral arms are a result of the interaction of the outer disc and the outer companion. In HD~100453 both the inner and outer disc will precess in a retrograde sense because the precession is caused by the binary \citep{Bate:2000fk}. However, recent work by \citet{Nealon:2020bq} has shown that shadows cast by strongly misaligned discs will rock back and forth in a restricted azimuthal range, even as the inner disc precession is retrograde. With a relative disc misalignment of $72 \degr$, these `rocking shadows' are expected in HD~100453. Thus if the shadows in HD~100453 are observed to move their direction of motion will not necessarily be in the same sense as the precession of the disc.

%-------------------------------------------------------------
%-------------------------------------------------------------

\section{Conclusion}
\label{section:conc}

The protoplanetary disc HD~100453 has an inner cavity, a strongly misaligned inner disc, an outer disc with two symmetric spirals and a bound outer companion. To successfully explain the origin of the features observed in the disc we found it necessary to consider all the components of the system. In our companion paper, \citetalias{Gonzalez:sub}, we showed that the orbit of HD~100453~B is misaligned by $61\degr$ to the plane of the outer disc in order to match the spiral features and velocity structure. In this work, we investigated the presence of an inner companion to explain the inner cavity and the inner misaligned disc.

We established that the observed misalignment between the inner and outer disc cannot be explained by the observed binary companion alone. This adds to the growing weight of evidence that there is an inner companion residing in the disc. Using numerical simulations, we showed that the planet is likely to have a mass lower than has been previously suggested. Our simulations also suggested that the planet is likely to be circulating, where the plane of its orbit is misaligned to that of the outer disc, with its evolution mostly governed by Kozai-Lidov oscillations. Due to the presence of the outer companion, such a planet would be difficult to detect in existing kinematics. Higher resolution kinematics or direct imaging may more clearly show the presence of the planet.

The misalignment between the inner and outer disc of HD~100453 is easily explained with the addition of this inner companion. In our proposed scenario the inner disc, inner companion and outer disc all start aligned. The outer companion is misaligned to this plane by $\sim 60 \degr$ and causes differential precession and potentially Kozai-Lidov oscillations that drive the outer disc to precess more rapidly than the inner one. This forms a relative misalignment between the inner and outer disc which is enhanced by ongoing differential precession from the outer companion. We refer to this scenario --- where the outer disc precesses more rapidly than the inner disc --- as `precession inception'. Shadows are expected to naturally arise in such a geometrical configuration. Therefore, it is only by consideration of all the components of the HD~100453 system that we can naturally explain the origin of the misalignments. While this scenario does hinge on the misalignment of the outer companion demonstrated in \citetalias{Gonzalez:sub}, there is currently no robust alternative scenario that results in such a large misalignment between the inner and outer disc.

It appears that HD~100453 is currently unique amongst protoplanetary discs that show evidence of disc misalignment. However, the precession inception mechanism could potentially be at work in other systems exhibiting similar disc features (e.g. cavity, spirals, shadows). The only requirements are two suitably disconnected discs (in this case by a companion) and the presence of a misaligned external torque. Once these conditions are met, differential precession means that misalignment is unavoidable and potentially long lived between the different components. Therefore, this scenario constitutes a new dynamical pathway to produce highly inclined circumstellar discs that cast shadows.

%It appears that this scenario is currently unique amongst protoplanetary discs that show evidence of disc misalignment. However, the only requirements we have for this scenario are that the discs be suitably disconnected (in this case by a companion) and the presence of a misaligned external torque. It is only by consideration of all the components of the HD~100453 system that we can naturally explain the origin of the misalignments.}

%-------------------------------------------------------------
%-------------------------------------------------------------

\section*{Acknowledgements}
The authors thank the referee for constructive comments. This project has received funding from the European Research Council (ERC) under the European Union's Horizon 2020 research and innovation programme (grant agreement No 681601). This project has received funding from the European Union's Horizon 2020 research and innovation programme under the Marie Sk\l{}odowska-Curie grant agreements No 210021 and No 823823 (DUSTBUSTERS). JFG, GvdP, and FMe acknowledge funding from ANR (Agence Nationale de la Recherche) of France under contract number ANR-16-CE31-0013 (Planet-Forming-Disks). JFG thanks the LABEX Lyon Institute of Origins (ANR-10-LABX-0066) of the Universit\'e de Lyon for its financial support within the programme `Investissements d'Avenir' (ANR-11-IDEX-0007) of the French government operated by the ANR. C.P. and D.J.P. acknowledge funding from the Australian Research Council via
FT170100040, FT130100034, and DP180104235. This paper makes use of the following ALMA data: ADS/JAO.ALMA\#2017.1.01424.S. ALMA is a partnership of ESO (representing its member states), NSF (USA) and NINS (Japan), together with NRC (Canada), MOST and ASIAA (Taiwan), and KASI (Republic of Korea), in cooperation with the Republic of Chile. The Joint ALMA Observatory is operated by ESO, AUI/NRAO and NAOJ. Figure~\ref{fig:render} was plotted with \textsc{splash} \citet{Price:2007kx} and Figure~\ref{fig:CO_map} with \textsc{pymcfost}. All other figures were produced using the community open-source Python package Matplotlib \citep{matplotlib}. This work was performed using the DiRAC Data Intensive service at Leicester, operated by the University of Leicester IT Services, which forms part of the STFC DiRAC HPC Facility (www.dirac.ac.uk). The equipment was funded by BEIS capital funding via STFC capital grants ST/K000373/1 and ST/R002363/1 and STFC DiRAC Operations grant ST/R001014/1. DiRAC is part of the National e-Infrastructure.

\section*{Data Availability Statement}
Hydrodynamic simulations used the \textsc{Phantom} code which is available from \url{https://github.com/danieljprice/phantom}. Radiative transfer calculations were made using \textsc{mcfost} which is available on a collaborative basis. N-body simulations in this paper made use of the \textsc{REBOUND} code which is freely available at \url{http://github.com/hannorein/rebound}. The input files for generating the SPH simulations, radiative transfer models and \textsc{REBOUND} calculations will be shared on reasonable request to the corresponding author.



\bibliographystyle{mnras}
\bibliography{precession_inception} % if your bibtex file is called example.bib

\begin{thebibliography}{}
\makeatletter
\relax
\def\mn@urlcharsother{\let\do\@makeother \do\$\do\&\do\#\do\^\do\_\do\%\do\~}
\def\mn@doi{\begingroup\mn@urlcharsother \@ifnextchar [ {\mn@doi@}
  {\mn@doi@[]}}
\def\mn@doi@[#1]#2{\def\@tempa{#1}\ifx\@tempa\@empty \href
  {http://dx.doi.org/#2} {doi:#2}\else \href {http://dx.doi.org/#2} {#1}\fi
  \endgroup}
\def\mn@eprint#1#2{\mn@eprint@#1:#2::\@nil}
\def\mn@eprint@arXiv#1{\href {http://arxiv.org/abs/#1} {{\tt arXiv:#1}}}
\def\mn@eprint@dblp#1{\href {http://dblp.uni-trier.de/rec/bibtex/#1.xml}
  {dblp:#1}}
\def\mn@eprint@#1:#2:#3:#4\@nil{\def\@tempa {#1}\def\@tempb {#2}\def\@tempc
  {#3}\ifx \@tempc \@empty \let \@tempc \@tempb \let \@tempb \@tempa \fi \ifx
  \@tempb \@empty \def\@tempb {arXiv}\fi \@ifundefined
  {mn@eprint@\@tempb}{\@tempb:\@tempc}{\expandafter \expandafter \csname
  mn@eprint@\@tempb\endcsname \expandafter{\@tempc}}}

\bibitem[\protect\citeauthoryear{{Bai} \& {Stone}}{{Bai} \&
  {Stone}}{2013}]{bai_stone_2012}
{Bai} X.-N.,  {Stone} J.~M.,  2013, \mn@doi [\apj]
  {10.1088/0004-637X/767/1/30}, \href
  {http://adsabs.harvard.edu/abs/2013ApJ...767...30B} {767, 30}

\bibitem[\protect\citeauthoryear{{Bate}}{{Bate}}{2018}]{Bate:2018ls}
{Bate} M.~R.,  2018, \mn@doi [\mnras] {10.1093/mnras/sty169}, \href
  {http://ukads.nottingham.ac.uk/abs/2018MNRAS.475.5618B} {475, 5618}

\bibitem[\protect\citeauthoryear{{Bate}, {Bonnell}, {Clarke}, {Lubow},
  {Ogilvie}, {Pringle}  \& {Tout}}{{Bate} et~al.}{2000}]{Bate:2000fk}
{Bate} M.~R.,  {Bonnell} I.~A.,  {Clarke} C.~J.,  {Lubow} S.~H.,  {Ogilvie}
  G.~I.,  {Pringle} J.~E.,   {Tout} C.~A.,  2000, \mn@doi [\mnras]
  {10.1046/j.1365-8711.2000.03648.x}, \href
  {http://adsabs.harvard.edu/abs/2000MNRAS.317..773B} {317, 773}

\bibitem[\protect\citeauthoryear{{Batygin}}{{Batygin}}{2012}]{Batygin:2012ge}
{Batygin} K.,  2012, \mn@doi [\nat] {10.1038/nature11560}, \href
  {https://ui.adsabs.harvard.edu/abs/2012Natur.491..418B} {491, 418}

\bibitem[\protect\citeauthoryear{{Batygin}, {Morbidelli}  \&
  {Tsiganis}}{{Batygin} et~al.}{2011}]{Batygin:2011oj}
{Batygin} K.,  {Morbidelli} A.,   {Tsiganis} K.,  2011, \mn@doi [\aap]
  {10.1051/0004-6361/201117193}, \href
  {https://ui.adsabs.harvard.edu/abs/2011A&A...533A...7B} {533, A7}

\bibitem[\protect\citeauthoryear{{Benisty} et~al.,}{{Benisty}
  et~al.}{2017}]{Benisty:2017kq}
{Benisty} M.,  et~al., 2017, \mn@doi [\aap] {10.1051/0004-6361/201629798},
  \href {http://adsabs.harvard.edu/abs/2017A%26A...597A..42B} {597, A42}

\bibitem[\protect\citeauthoryear{{Benisty} et~al.,}{{Benisty}
  et~al.}{2018}]{Benisty:2018ve}
{Benisty} M.,  et~al., 2018, \mn@doi [\aap] {10.1051/0004-6361/201833913},
  \href {https://ui.adsabs.harvard.edu/abs/2018A&A...619A.171B} {619, A171}

\bibitem[\protect\citeauthoryear{{B{\'e}thune}, {Lesur}  \&
  {Ferreira}}{{B{\'e}thune} et~al.}{2016}]{Bethune:2016wf}
{B{\'e}thune} W.,  {Lesur} G.,   {Ferreira} J.,  2016, \mn@doi [\aap]
  {10.1051/0004-6361/201527874}, \href
  {https://ui.adsabs.harvard.edu/abs/2016A&A...589A..87B} {589, A87}

\bibitem[\protect\citeauthoryear{{Bitsch}, {Crida}, {Libert}  \&
  {Lega}}{{Bitsch} et~al.}{2013}]{Bitsch:2013hg}
{Bitsch} B.,  {Crida} A.,  {Libert} A.-S.,   {Lega} E.,  2013, \mn@doi [\aap]
  {10.1051/0004-6361/201220310}, \href
  {http://adsabs.harvard.edu/abs/2013A%26A...555A.124B} {555, A124}

\bibitem[\protect\citeauthoryear{{Boccaletti} et~al.,}{{Boccaletti}
  et~al.}{2020}]{Boccaletti:2020sp}
{Boccaletti} A.,  et~al., 2020, \mn@doi [\aap] {10.1051/0004-6361/202038008},
  \href {https://ui.adsabs.harvard.edu/abs/2020A&A...637L...5B} {637, L5}

\bibitem[\protect\citeauthoryear{{Casassus} et~al.,}{{Casassus}
  et~al.}{2015}]{Casassus:2015yu}
{Casassus} S.,  et~al., 2015, \mn@doi [\apj] {10.1088/0004-637X/811/2/92},
  \href {http://adsabs.harvard.edu/abs/2015ApJ...811...92C} {811, 92}

\bibitem[\protect\citeauthoryear{{Chen}, {Henning}, {van Boekel}  \&
  {Grady}}{{Chen} et~al.}{2006}]{Chen:2006vs}
{Chen} X.~P.,  {Henning} T.,  {van Boekel} R.,   {Grady} C.~A.,  2006, \mn@doi
  [\aap] {10.1051/0004-6361:20054122}, \href
  {https://ui.adsabs.harvard.edu/abs/2006A&A...445..331C} {445, 331}

\bibitem[\protect\citeauthoryear{{Collins} et~al.,}{{Collins}
  et~al.}{2009}]{Collins:2009he}
{Collins} K.~A.,  et~al., 2009, \mn@doi [\apj] {10.1088/0004-637X/697/1/557},
  \href {https://ui.adsabs.harvard.edu/abs/2009ApJ...697..557C} {697, 557}

\bibitem[\protect\citeauthoryear{{Crida}, {Morbidelli}  \& {Masset}}{{Crida}
  et~al.}{2006}]{Crida:2006bv}
{Crida} A.,  {Morbidelli} A.,   {Masset} F.,  2006, \mn@doi [\icarus]
  {10.1016/j.icarus.2005.10.007}, \href
  {https://ui.adsabs.harvard.edu/#abs/2006Icar..181..587C} {181, 587}

\bibitem[\protect\citeauthoryear{{Cuello} et~al.,}{{Cuello}
  et~al.}{2019}]{Cuello:2019bd}
{Cuello} N.,  et~al., 2019, \mn@doi [\mnras] {10.1093/mnras/sty3325}, \href
  {https://ui.adsabs.harvard.edu/abs/2019MNRAS.483.4114C} {483, 4114}

\bibitem[\protect\citeauthoryear{{Cuello} et~al.,}{{Cuello}
  et~al.}{2020}]{Cuello:2020rt}
{Cuello} N.,  et~al., 2020, \mn@doi [\mnras] {10.1093/mnras/stz2938}, \href
  {https://ui.adsabs.harvard.edu/abs/2020MNRAS.491..504C} {491, 504}

\bibitem[\protect\citeauthoryear{{Dipierro} \& {Laibe}}{{Dipierro} \&
  {Laibe}}{2017}]{Dipierro:2017kq}
{Dipierro} G.,  {Laibe} G.,  2017, \mn@doi [\mnras] {10.1093/mnras/stx977},
  \href {https://ui.adsabs.harvard.edu/abs/2017MNRAS.469.1932D} {469, 1932}

\bibitem[\protect\citeauthoryear{{Dipierro}, {Price}, {Laibe}, {Hirsh},
  {Cerioli}  \& {Lodato}}{{Dipierro} et~al.}{2015}]{Dipierro:2015od}
{Dipierro} G.,  {Price} D.,  {Laibe} G.,  {Hirsh} K.,  {Cerioli} A.,   {Lodato}
  G.,  2015, \mn@doi [\mnras] {10.1093/mnrasl/slv105}, \href
  {http://adsabs.harvard.edu/abs/2015MNRAS.453L..73D} {453, L73}

\bibitem[\protect\citeauthoryear{{Dipierro}, {Laibe}, {Alexander}  \&
  {Hutchison}}{{Dipierro} et~al.}{2018}]{Dipierro:2018qp}
{Dipierro} G.,  {Laibe} G.,  {Alexander} R.,   {Hutchison} M.,  2018, \mn@doi
  [\mnras] {10.1093/mnras/sty1701}, \href
  {https://ui.adsabs.harvard.edu/abs/2018MNRAS.479.4187D} {479, 4187}

\bibitem[\protect\citeauthoryear{{Dong}, {Zhu}, {Fung}, {Rafikov}, {Chiang}  \&
  {Wagner}}{{Dong} et~al.}{2016}]{Dong:2016oc}
{Dong} R.,  {Zhu} Z.,  {Fung} J.,  {Rafikov} R.,  {Chiang} E.,   {Wagner} K.,
  2016, \mn@doi [\apjl] {10.3847/2041-8205/816/1/L12}, \href
  {http://adsabs.harvard.edu/abs/2016ApJ...816L..12D} {816, L12}

\bibitem[\protect\citeauthoryear{{Do{\v{g}}an}, {Nixon}, {King}  \&
  {Pringle}}{{Do{\v{g}}an} et~al.}{2018}]{Dogan:2018wo}
{Do{\v{g}}an} S.,  {Nixon} C.~J.,  {King} A.~R.,   {Pringle} J.~E.,  2018,
  \mn@doi [\mnras] {10.1093/mnras/sty155}, \href
  {https://ui.adsabs.harvard.edu/abs/2018MNRAS.476.1519D} {476, 1519}

\bibitem[\protect\citeauthoryear{{Duffell} \& {MacFadyen}}{{Duffell} \&
  {MacFadyen}}{2013}]{Duffel:2013mu}
{Duffell} P.~C.,  {MacFadyen} A.~I.,  2013, \mn@doi [\apj]
  {10.1088/0004-637X/769/1/41}, \href
  {https://ui.adsabs.harvard.edu/abs/2013ApJ...769...41D} {769, 41}

\bibitem[\protect\citeauthoryear{{Facchini}, {Lodato}  \& {Price}}{{Facchini}
  et~al.}{2013}]{facchini_2013}
{Facchini} S.,  {Lodato} G.,   {Price} D.~J.,  2013, \mn@doi [\mnras]
  {10.1093/mnras/stt877}, \href
  {http://adsabs.harvard.edu/abs/2013MNRAS.433.2142F} {433, 2142}

\bibitem[\protect\citeauthoryear{{Facchini}, {Juh{\'a}sz}  \&
  {Lodato}}{{Facchini} et~al.}{2018}]{Facchini:2017of}
{Facchini} S.,  {Juh{\'a}sz} A.,   {Lodato} G.,  2018, \mn@doi [\mnras]
  {10.1093/mnras/stx2523}, \href
  {https://ui.adsabs.harvard.edu/\#abs/2018MNRAS.473.4459F} {473, 4459}

\bibitem[\protect\citeauthoryear{{Flaherty}, {Hughes}, {Rosenfeld}, {Andrews},
  {Chiang}, {Simon}, {Kerzner}  \& {Wilner}}{{Flaherty}
  et~al.}{2015}]{Flaherty:2015is}
{Flaherty} K.~M.,  {Hughes} A.~M.,  {Rosenfeld} K.~A.,  {Andrews} S.~M.,
  {Chiang} E.,  {Simon} J.~B.,  {Kerzner} S.,   {Wilner} D.~J.,  2015, \mn@doi
  [\apj] {10.1088/0004-637X/813/2/99}, \href
  {http://ukads.nottingham.ac.uk/abs/2015ApJ...813...99F} {813, 99}

\bibitem[\protect\citeauthoryear{{Flaherty} et~al.,}{{Flaherty}
  et~al.}{2017}]{Flaherty:2017do}
{Flaherty} K.~M.,  et~al., 2017, \mn@doi [\apj] {10.3847/1538-4357/aa79f9},
  \href {http://ukads.nottingham.ac.uk/abs/2017ApJ...843..150F} {843, 150}

\bibitem[\protect\citeauthoryear{{Flock}, {Ruge}, {Dzyurkevich}, {Henning},
  {Klahr}  \& {Wolf}}{{Flock} et~al.}{2015}]{Flock:2015vq}
{Flock} M.,  {Ruge} J.~P.,  {Dzyurkevich} N.,  {Henning} T.,  {Klahr} H.,
  {Wolf} S.,  2015, \mn@doi [\aap] {10.1051/0004-6361/201424693}, \href
  {https://ui.adsabs.harvard.edu/abs/2015A&A...574A..68F} {574, A68}

\bibitem[\protect\citeauthoryear{{Franchini}, {Martin}  \& {Lubow}}{{Franchini}
  et~al.}{2019}]{Franchini:2019gc}
{Franchini} A.,  {Martin} R.~G.,   {Lubow} S.~H.,  2019, \mn@doi [\mnras]
  {10.1093/mnras/stz424}, \href
  {https://ui.adsabs.harvard.edu/abs/2019MNRAS.485..315F} {485, 315}

\bibitem[\protect\citeauthoryear{{Franchini}, {Martin}  \& {Lubow}}{{Franchini}
  et~al.}{2020}]{Franchini:2020ht}
{Franchini} A.,  {Martin} R.~G.,   {Lubow} S.~H.,  2020, \mn@doi [\mnras]
  {10.1093/mnras/stz3175}, \href
  {https://ui.adsabs.harvard.edu/abs/2020MNRAS.491.5351F} {491, 5351}

\bibitem[\protect\citeauthoryear{{Fu}, {Lubow}  \& {Martin}}{{Fu}
  et~al.}{2015}]{Fu:2015wg}
{Fu} W.,  {Lubow} S.~H.,   {Martin} R.~G.,  2015, \mn@doi [\apj]
  {10.1088/0004-637X/813/2/105}, \href
  {https://ui.adsabs.harvard.edu/abs/2015ApJ...813..105F} {813, 105}

\bibitem[\protect\citeauthoryear{{Gonzalez}, {Laibe}  \& {Maddison}}{{Gonzalez}
  et~al.}{2017}]{Gonzalez:2017bu}
{Gonzalez} J.~F.,  {Laibe} G.,   {Maddison} S.~T.,  2017, \mn@doi [Monthly
  Notices of the Royal Astronomical Society] {10.1093/mnras/stx016}, \href
  {https://ui.adsabs.harvard.edu/abs/2017MNRAS.467.1984G} {467, 1984}

\bibitem[\protect\citeauthoryear{{Gonzalez} et~al.,}{{Gonzalez}
  et~al.}{2020}]{Gonzalez:sub}
{Gonzalez} J.-F.,  et~al., 2020, \mnras, submitted (Paper~I)

\bibitem[\protect\citeauthoryear{Hunter}{Hunter}{2007}]{matplotlib}
Hunter J.~D.,  2007, \mn@doi [Computing In Science \& Engineering]
  {10.1109/MCSE.2007.55}, 9, 90

\bibitem[\protect\citeauthoryear{{Keppler} et~al.,}{{Keppler}
  et~al.}{2018}]{Keppler:2018pd}
{Keppler} M.,  et~al., 2018, \mn@doi [\aap] {10.1051/0004-6361/201832957},
  \href {https://ui.adsabs.harvard.edu/abs/2018A&A...617A..44K} {617, A44}

\bibitem[\protect\citeauthoryear{{Keppler} et~al.,}{{Keppler}
  et~al.}{2019}]{Keppler:2019ce}
{Keppler} M.,  et~al., 2019, \mn@doi [\aap] {10.1051/0004-6361/201935034},
  \href {https://ui.adsabs.harvard.edu/abs/2019A&A...625A.118K} {625, A118}

\bibitem[\protect\citeauthoryear{{Kluska} et~al.,}{{Kluska}
  et~al.}{2020}]{Kluska:2020og}
{Kluska} J.,  et~al., 2020, \mn@doi [\aap] {10.1051/0004-6361/201833774}, \href
  {https://ui.adsabs.harvard.edu/abs/2020A&A...636A.116K} {636, A116}

\bibitem[\protect\citeauthoryear{{Kozai}}{{Kozai}}{1962}]{Kozai:1962gq}
{Kozai} Y.,  1962, \mn@doi [\aj] {10.1086/108790}, \href
  {https://ui.adsabs.harvard.edu/abs/1962AJ.....67..591K} {67, 591}

\bibitem[\protect\citeauthoryear{{Lazareff} et~al.,}{{Lazareff}
  et~al.}{2017}]{Lazareff:2017ug}
{Lazareff} B.,  et~al., 2017, \mn@doi [\aap] {10.1051/0004-6361/201629305},
  \href {https://ui.adsabs.harvard.edu/abs/2017A&A...599A..85L} {599, A85}

\bibitem[\protect\citeauthoryear{{Li}, {Naoz}, {Holman}  \& {Loeb}}{{Li}
  et~al.}{2014}]{Li:2014ua}
{Li} G.,  {Naoz} S.,  {Holman} M.,   {Loeb} A.,  2014, \mn@doi [\apj]
  {10.1088/0004-637X/791/2/86}, \href
  {https://ui.adsabs.harvard.edu/abs/2014ApJ...791...86L} {791, 86}

\bibitem[\protect\citeauthoryear{{Lidov}}{{Lidov}}{1962}]{Lidov:1962gq}
{Lidov} M.~L.,  1962, \mn@doi [\planss] {10.1016/0032-0633(62)90129-0}, \href
  {https://ui.adsabs.harvard.edu/abs/1962P&SS....9..719L} {9, 719}

\bibitem[\protect\citeauthoryear{{Long} et~al.,}{{Long}
  et~al.}{2018}]{Long:2018vt}
{Long} F.,  et~al., 2018, \mn@doi [The Astrophysical Journal]
  {10.3847/1538-4357/aae8e1}, \href
  {https://ui.adsabs.harvard.edu/abs/2018ApJ...869...17L} {869, 17}

\bibitem[\protect\citeauthoryear{{Lubow} \& {Martin}}{{Lubow} \&
  {Martin}}{2016}]{Lubow:2016nw}
{Lubow} S.~H.,  {Martin} R.~G.,  2016, \mn@doi [\apj]
  {10.3847/0004-637X/817/1/30}, \href
  {https://ui.adsabs.harvard.edu/#abs/2016ApJ...817...30L} {817, 30}

\bibitem[\protect\citeauthoryear{{Lubow} \& {Ogilvie}}{{Lubow} \&
  {Ogilvie}}{2017}]{Lubow:2017hw}
{Lubow} S.~H.,  {Ogilvie} G.~I.,  2017, \mn@doi [\mnras]
  {10.1093/mnras/stx990}, \href
  {https://ui.adsabs.harvard.edu/abs/2017MNRAS.469.4292L} {469, 4292}

\bibitem[\protect\citeauthoryear{{Marino}, {Perez}  \& {Casassus}}{{Marino}
  et~al.}{2015}]{Marino:2015rh}
{Marino} S.,  {Perez} S.,   {Casassus} S.,  2015, \mn@doi [\apjl]
  {10.1088/2041-8205/798/2/L44}, \href
  {http://adsabs.harvard.edu/abs/2015ApJ...798L..44M} {798, L44}

\bibitem[\protect\citeauthoryear{{Martin}, {Nixon}, {Lubow}, {Armitage},
  {Price}, {Do{\u g}an}  \& {King}}{{Martin} et~al.}{2014}]{Martin:2014bb}
{Martin} R.~G.,  {Nixon} C.,  {Lubow} S.~H.,  {Armitage} P.~J.,  {Price} D.~J.,
   {Do{\u g}an} S.,   {King} A.,  2014, \mn@doi [\apjl]
  {10.1088/2041-8205/792/2/L33}, \href
  {http://adsabs.harvard.edu/abs/2014ApJ...792L..33M} {792, L33}

\bibitem[\protect\citeauthoryear{{Martin}, {Lubow}, {Nixon}  \&
  {Armitage}}{{Martin} et~al.}{2016}]{Martin:2016qf}
{Martin} R.~G.,  {Lubow} S.~H.,  {Nixon} C.,   {Armitage} P.~J.,  2016, \mn@doi
  [\mnras] {10.1093/mnras/stw605}, \href
  {http://adsabs.harvard.edu/abs/2016MNRAS.458.4345M} {458, 4345}

\bibitem[\protect\citeauthoryear{{M{\'e}nard} et~al.,}{{M{\'e}nard}
  et~al.}{2020}]{Menard:2020aa}
{M{\'e}nard} F.,  et~al., 2020, \mn@doi [\aap] {10.1051/0004-6361/202038356},
  \href {https://ui.adsabs.harvard.edu/abs/2020A&A...639L...1M} {639, L1}

\bibitem[\protect\citeauthoryear{{Menu}, {van Boekel}, {Henning}, {Leinert},
  {Waelkens}  \& {Waters}}{{Menu} et~al.}{2015}]{Menu:2015vy}
{Menu} J.,  {van Boekel} R.,  {Henning} T.,  {Leinert} C.,  {Waelkens} C.,
  {Waters} L.~B.~F.~M.,  2015, \mn@doi [\aap] {10.1051/0004-6361/201525654},
  \href {https://ui.adsabs.harvard.edu/abs/2015A&A...581A.107M} {581, A107}

\bibitem[\protect\citeauthoryear{{Min}, {Stolker}, {Dominik}  \&
  {Benisty}}{{Min} et~al.}{2017}]{Min:2017oc}
{Min} M.,  {Stolker} T.,  {Dominik} C.,   {Benisty} M.,  2017, \mn@doi [\aap]
  {10.1051/0004-6361/201730949}, \href
  {http://ukads.nottingham.ac.uk/abs/2017A%26A...604L..10M} {604, L10}

\bibitem[\protect\citeauthoryear{{Montesinos} \& {Cuello}}{{Montesinos} \&
  {Cuello}}{2018}]{Montesinos:2018zd}
{Montesinos} M.,  {Cuello} N.,  2018, \mn@doi [\mnras] {10.1093/mnrasl/sly001},
  \href {https://ui.adsabs.harvard.edu/abs/2018MNRAS.475L..35M} {475, L35}

\bibitem[\protect\citeauthoryear{{Montesinos}, {Perez}, {Casassus}, {Marino},
  {Cuadra}  \& {Christiaens}}{{Montesinos} et~al.}{2016}]{Montesinos:2016ig}
{Montesinos} M.,  {Perez} S.,  {Casassus} S.,  {Marino} S.,  {Cuadra} J.,
  {Christiaens} V.,  2016, \mn@doi [\apj] {10.3847/2041-8205/823/1/L8}, \href
  {https://ui.adsabs.harvard.edu/abs/2016ApJ...823L...8M} {823, L8}

\bibitem[\protect\citeauthoryear{{M{\"u}ller} et~al.,}{{M{\"u}ller}
  et~al.}{2018}]{Mueller:2018vw}
{M{\"u}ller} A.,  et~al., 2018, \mn@doi [\aap] {10.1051/0004-6361/201833584},
  \href {https://ui.adsabs.harvard.edu/abs/2018A&A...617L...2M} {617, L2}

\bibitem[\protect\citeauthoryear{{Nealon}, {Dipierro}, {Alexander}, {Martin}
  \& {Nixon}}{{Nealon} et~al.}{2018}]{Nealon:2018ic}
{Nealon} R.,  {Dipierro} G.,  {Alexander} R.,  {Martin} R.~G.,   {Nixon} C.,
  2018, \mn@doi [\mnras] {10.1093/mnras/sty2267}, \href
  {https://ui.adsabs.harvard.edu/\#abs/2018MNRAS.481...20N} {481, 20}

\bibitem[\protect\citeauthoryear{{Nealon}, {Price}  \& {Pinte}}{{Nealon}
  et~al.}{2020}]{Nealon:2020bq}
{Nealon} R.,  {Price} D.~J.,   {Pinte} C.,  2020, \mn@doi [\mnras]
  {10.1093/mnrasl/slaa026}, \href
  {https://ui.adsabs.harvard.edu/abs/2020MNRAS.493L.143N} {493, L143}

\bibitem[\protect\citeauthoryear{{Nixon}, {King}  \& {Price}}{{Nixon}
  et~al.}{2013}]{nixon_2013}
{Nixon} C.,  {King} A.,   {Price} D.,  2013, \mn@doi [\mnras]
  {10.1093/mnras/stt1136}, \href
  {http://adsabs.harvard.edu/abs/2013MNRAS.434.1946N} {434, 1946}

\bibitem[\protect\citeauthoryear{{Okuzumi}, {Momose}, {Sirono}, {Kobayashi}  \&
  {Tanaka}}{{Okuzumi} et~al.}{2016}]{Okuzumi:2016by}
{Okuzumi} S.,  {Momose} M.,  {Sirono} S.-i.,  {Kobayashi} H.,   {Tanaka} H.,
  2016, \mn@doi [The Astrophysical Journal] {10.3847/0004-637X/821/2/82}, \href
  {https://ui.adsabs.harvard.edu/abs/2016ApJ...821...82O} {821, 82}

\bibitem[\protect\citeauthoryear{{Owen} \& {Lai}}{{Owen} \&
  {Lai}}{2017}]{Owen:2017oj}
{Owen} J.~E.,  {Lai} D.,  2017, \mn@doi [\mnras] {10.1093/mnras/stx1033}, \href
  {https://ui.adsabs.harvard.edu/abs/2017MNRAS.469.2834O} {469, 2834}

\bibitem[\protect\citeauthoryear{{Picogna} \& {Marzari}}{{Picogna} \&
  {Marzari}}{2015}]{Picogna:2015bo}
{Picogna} G.,  {Marzari} F.,  2015, \mn@doi [\aap]
  {10.1051/0004-6361/201526162}, \href
  {https://ui.adsabs.harvard.edu/abs/2015A&A...583A.133P} {583, A133}

\bibitem[\protect\citeauthoryear{{Pinilla} et~al.,}{{Pinilla}
  et~al.}{2018}]{Pinilla:2018gb}
{Pinilla} P.,  et~al., 2018, \mn@doi [\apj] {10.3847/1538-4357/aae824}, \href
  {https://ui.adsabs.harvard.edu/abs/2018ApJ...868...85P} {868, 85}

\bibitem[\protect\citeauthoryear{{Pinte}, {M{\'e}nard}, {Duch{\^e}ne}  \&
  {Bastien}}{{Pinte} et~al.}{2006}]{Pinte:2006nw}
{Pinte} C.,  {M{\'e}nard} F.,  {Duch{\^e}ne} G.,   {Bastien} P.,  2006, \mn@doi
  [\aap] {10.1051/0004-6361:20053275}, \href
  {http://adsabs.harvard.edu/abs/2006A%26A...459..797P} {459, 797}

\bibitem[\protect\citeauthoryear{{Pinte}, {Harries}, {Min}, {Watson},
  {Dullemond}, {Woitke}, {M{\'e}nard}  \& {Dur{\'a}n-Rojas}}{{Pinte}
  et~al.}{2009}]{Pinte:2009ye}
{Pinte} C.,  {Harries} T.~J.,  {Min} M.,  {Watson} A.~M.,  {Dullemond} C.~P.,
  {Woitke} P.,  {M{\'e}nard} F.,   {Dur{\'a}n-Rojas} M.~C.,  2009, \mn@doi
  [\aap] {10.1051/0004-6361/200811555}, \href
  {http://adsabs.harvard.edu/abs/2009A%26A...498..967P} {498, 967}

\bibitem[\protect\citeauthoryear{{Pinte}, {Dent}, {M{\'e}nard}, {Hales},
  {Hill}, {Cortes}  \& {de Gregorio-Monsalvo}}{{Pinte}
  et~al.}{2016}]{Pinte:2016oa}
{Pinte} C.,  {Dent} W.~R.~F.,  {M{\'e}nard} F.,  {Hales} A.,  {Hill} T.,
  {Cortes} P.,   {de Gregorio-Monsalvo} I.,  2016, \mn@doi [\apj]
  {10.3847/0004-637X/816/1/25}, \href
  {http://ukads.nottingham.ac.uk/abs/2016ApJ...816...25P} {816, 25}

\bibitem[\protect\citeauthoryear{{Pinte} et~al.,}{{Pinte}
  et~al.}{2018}]{Pinte:2018ah}
{Pinte} C.,  et~al., 2018, \mn@doi [\apjl] {10.3847/2041-8213/aac6dc}, \href
  {https://ui.adsabs.harvard.edu/abs/2018ApJ...860L..13P} {860, L13}

\bibitem[\protect\citeauthoryear{{Pinte} et~al.,}{{Pinte}
  et~al.}{2019}]{Pinte:2019xo}
{Pinte} C.,  et~al., 2019, \mn@doi [Nature Astronomy]
  {10.1038/s41550-019-0852-6}, \href
  {https://ui.adsabs.harvard.edu/abs/2019NatAs...3.1109P} {3, 1109}

\bibitem[\protect\citeauthoryear{{Pinte} et~al.,}{{Pinte}
  et~al.}{2020}]{Pinte:2020gh}
{Pinte} C.,  et~al., 2020, \mn@doi [\apjl] {10.3847/2041-8213/ab6dda}, \href
  {https://ui.adsabs.harvard.edu/abs/2020ApJ...890L...9P} {890, L9}

\bibitem[\protect\citeauthoryear{{Poblete}, {Calcino}, {Cuello}, {Mac{\'\i}as},
  {Ribas}, {Price}, {Cuadra}  \& {Pinte}}{{Poblete}
  et~al.}{2020}]{Poblete:2020}
{Poblete} P.~P.,  {Calcino} J.,  {Cuello} N.,  {Mac{\'\i}as} E.,  {Ribas}
  {\'A}.,  {Price} D.~J.,  {Cuadra} J.,   {Pinte} C.,  2020, \mn@doi [\mnras]
  {10.1093/mnras/staa1655}, \href
  {https://ui.adsabs.harvard.edu/abs/2020MNRAS.496.2362P} {496, 2362}

\bibitem[\protect\citeauthoryear{{Price}}{{Price}}{2007}]{Price:2007kx}
{Price} D.~J.,  2007, \mn@doi [\pasa] {10.1071/AS07022}, \href
  {http://adsabs.harvard.edu/abs/2007PASA...24..159P} {24, 159}

\bibitem[\protect\citeauthoryear{{Price} et~al.,}{{Price}
  et~al.}{2018a}]{Phantom}
{Price} D.~J.,  et~al., 2018a, \mn@doi [\pasa] {10.1017/pasa.2018.25}, \href
  {https://ui.adsabs.harvard.edu/\#abs/2018PASA...35...31P} {35, e031}

\bibitem[\protect\citeauthoryear{{Price} et~al.,}{{Price}
  et~al.}{2018b}]{Price:2018pf}
{Price} D.~J.,  et~al., 2018b, \mn@doi [\mnras] {10.1093/mnras/sty647}, \href
  {http://ukads.nottingham.ac.uk/abs/2018MNRAS.477.1270P} {477, 1270}

\bibitem[\protect\citeauthoryear{{Rein}}{{Rein}}{2012}]{Rein:2012so}
{Rein} H.,  2012, \mn@doi [\mnras] {10.1111/j.1365-2966.2012.20869.x}, \href
  {http://ukads.nottingham.ac.uk/abs/2012MNRAS.422.3611R} {422, 3611}

\bibitem[\protect\citeauthoryear{{Rein} \& {Liu}}{{Rein} \&
  {Liu}}{2012}]{REBOUND}
{Rein} H.,  {Liu} S.~F.,  2012, \mn@doi [\aap] {10.1051/0004-6361/201118085},
  \href {https://ui.adsabs.harvard.edu/abs/2012A&A...537A.128R} {537, A128}

\bibitem[\protect\citeauthoryear{{Rein} \& {Spiegel}}{{Rein} \&
  {Spiegel}}{2015}]{Rein:2015nj}
{Rein} H.,  {Spiegel} D.~S.,  2015, \mn@doi [\mnras] {10.1093/mnras/stu2164},
  \href {https://ui.adsabs.harvard.edu/abs/2015MNRAS.446.1424R} {446, 1424}

\bibitem[\protect\citeauthoryear{{Riols} \& {Lesur}}{{Riols} \&
  {Lesur}}{2019}]{Riols:2019cw}
{Riols} A.,  {Lesur} G.,  2019, \mn@doi [\aap] {10.1051/0004-6361/201834813},
  \href {https://ui.adsabs.harvard.edu/abs/2019A&A...625A.108R} {625, A108}

\bibitem[\protect\citeauthoryear{{Rosotti} et~al.,}{{Rosotti}
  et~al.}{2020}]{Rosotti:2020nj}
{Rosotti} G.~P.,  et~al., 2020, \mn@doi [\mnras] {10.1093/mnras/stz3090}, \href
  {https://ui.adsabs.harvard.edu/abs/2020MNRAS.491.1335R} {491, 1335}

\bibitem[\protect\citeauthoryear{{Ru{\'{\i}}z-Rodr{\'{\i}}guez}, {Ireland},
  {Cieza}  \& {Kraus}}{{Ru{\'{\i}}z-Rodr{\'{\i}}guez}
  et~al.}{2016}]{Ruiz:2016ne}
{Ru{\'{\i}}z-Rodr{\'{\i}}guez} D.,  {Ireland} M.,  {Cieza} L.,   {Kraus} A.,
  2016, \mn@doi [\mnras] {10.1093/mnras/stw2297}, \href
  {http://adsabs.harvard.edu/abs/2016MNRAS.463.3829R} {463, 3829}

\bibitem[\protect\citeauthoryear{{Sicilia-Aguilar}, {Manara}, {de Boer},
  {Benisty}, {Pinilla}  \& {Bouvier}}{{Sicilia-Aguilar}
  et~al.}{2020}]{Sicilia-Aguilar:2020xx}
{Sicilia-Aguilar} A.,  {Manara} C.~F.,  {de Boer} J.,  {Benisty} M.,  {Pinilla}
  P.,   {Bouvier} J.,  2020, \mn@doi [\aap] {10.1051/0004-6361/201936565},
  \href {https://ui.adsabs.harvard.edu/abs/2020A&A...633A..37S} {633, A37}

\bibitem[\protect\citeauthoryear{{Stammler}, {Birnstiel}, {Pani{\'c}},
  {Dullemond}  \& {Dominik}}{{Stammler} et~al.}{2017}]{Stammler:2017bw}
{Stammler} S.~M.,  {Birnstiel} T.,  {Pani{\'c}} O.,  {Dullemond} C.~P.,
  {Dominik} C.,  2017, \mn@doi [Astronomy and Astrophysics]
  {10.1051/0004-6361/201629041}, \href
  {https://ui.adsabs.harvard.edu/abs/2017A&A...600A.140S} {600, A140}

\bibitem[\protect\citeauthoryear{{Tanaka} \& {Ward}}{{Tanaka} \&
  {Ward}}{2004}]{Tanaka:2004od}
{Tanaka} H.,  {Ward} W.~R.,  2004, \mn@doi [\apj] {10.1086/380992}, \href
  {http://ukads.nottingham.ac.uk/abs/2004ApJ...602..388T} {602, 388}

\bibitem[\protect\citeauthoryear{{Teague} et~al.,}{{Teague}
  et~al.}{2018}]{Teague:2018vw}
{Teague} R.,  et~al., 2018, \mn@doi [\apj] {10.3847/1538-4357/aad80e}, \href
  {https://ui.adsabs.harvard.edu/#abs/2018ApJ...864..133T} {864, 133}

\bibitem[\protect\citeauthoryear{{Vioque}, {Oudmaijer}, {Baines},
  {Mendigut{\'\i}a}  \& {P{\'e}rez-Mart{\'\i}nez}}{{Vioque}
  et~al.}{2018}]{Vioque:2018pg}
{Vioque} M.,  {Oudmaijer} R.~D.,  {Baines} D.,  {Mendigut{\'\i}a} I.,
  {P{\'e}rez-Mart{\'\i}nez} R.,  2018, \mn@doi [\aap]
  {10.1051/0004-6361/201832870}, \href
  {https://ui.adsabs.harvard.edu/abs/2018A&A...620A.128V} {620, A128}

\bibitem[\protect\citeauthoryear{{Wagner}, {Apai}, {Kasper}  \&
  {Robberto}}{{Wagner} et~al.}{2015}]{Wagner:2015ph}
{Wagner} K.,  {Apai} D.,  {Kasper} M.,   {Robberto} M.,  2015, \mn@doi [\apjl]
  {10.1088/2041-8205/813/1/L2}, \href
  {https://ui.adsabs.harvard.edu/abs/2015ApJ...813L...2W} {813, L2}

\bibitem[\protect\citeauthoryear{{Wagner} et~al.,}{{Wagner}
  et~al.}{2018}]{Wagner:2018hs}
{Wagner} K.,  et~al., 2018, \mn@doi [\apj] {10.3847/1538-4357/aaa767}, \href
  {https://ui.adsabs.harvard.edu/abs/2018ApJ...854..130W} {854, 130}

\bibitem[\protect\citeauthoryear{{Wurster}, {Bate}  \& {Price}}{{Wurster}
  et~al.}{2019}]{Wurster:2019pb}
{Wurster} J.,  {Bate} M.~R.,   {Price} D.~J.,  2019, \mn@doi [\mnras]
  {10.1093/mnras/stz2215}, \href
  {https://ui.adsabs.harvard.edu/abs/2019MNRAS.489.1719W} {489, 1719}

\bibitem[\protect\citeauthoryear{{Xiang-Gruess} \& {Papaloizou}}{{Xiang-Gruess}
  \& {Papaloizou}}{2013}]{Xiang-Gruess:2013fg}
{Xiang-Gruess} M.,  {Papaloizou} J.~C.~B.,  2013, \mn@doi [\mnras]
  {10.1093/mnras/stt254}, \href
  {http://ukads.nottingham.ac.uk/abs/2013MNRAS.431.1320X} {431, 1320}

\bibitem[\protect\citeauthoryear{{Xiang-Gruess} \& {Papaloizou}}{{Xiang-Gruess}
  \& {Papaloizou}}{2014}]{Xiang-Gruess:2014ne}
{Xiang-Gruess} M.,  {Papaloizou} J.~C.~B.,  2014, \mn@doi [\mnras]
  {10.1093/mnras/stu308}, \href
  {http://adsabs.harvard.edu/abs/2014MNRAS.440.1179X} {440, 1179}

\bibitem[\protect\citeauthoryear{{Zhu}}{{Zhu}}{2019}]{Zhu:2018vf}
{Zhu} Z.,  2019, \mn@doi [\mnras] {10.1093/mnras/sty3358}, \href
  {https://ui.adsabs.harvard.edu/abs/2019MNRAS.483.4221Z} {483, 4221}

\bibitem[\protect\citeauthoryear{{van der Plas} et~al.,}{{van der Plas}
  et~al.}{2019}]{vanderPlas:2019gy}
{van der Plas} G.,  et~al., 2019, \mn@doi [\aap] {10.1051/0004-6361/201834134},
  \href {https://ui.adsabs.harvard.edu/abs/2019A&A...624A..33V} {624, A33}

\makeatother
\end{thebibliography}

%%%%%%%%%%%%%%%%%%%%%%%%%%%%%%%%%%%%%%%%%%%%%%%%%%

%%%%%%%%%%%%%%%%% APPENDICES %%%%%%%%%%%%%%%%%%%%%

\appendix

%%%%%%%%%%%%%%%%%%%%%%%%%%%%%%%%%%%%%%%%%%%%%%%%%%

% Don't change these lines
\bsp	% typesetting comment
\label{lastpage}
\end{document}